\RequirePackage{tikz}

\documentclass[pdflatex,sn-mathphys,iicol]{sn-jnl}

\usepackage{pgfplots}
\usepackage{subcaption}

\usepackage[capitalise]{cleveref}

\usepackage{xspace}

\usepackage{tikz}
\usetikzlibrary{calc}

\definecolor{color1}{rgb}{0.12156862745098039, 0.4666666666666667, 0.7058823529411765}
\definecolor{color2}{rgb}{1.0, 0.4980392156862745, 0.054901960784313725}
\definecolor{color3}{rgb}{0.17254901960784313, 0.6274509803921569, 0.17254901960784313}
\definecolor{color4}{rgb}{0.8392156862745098, 0.15294117647058825, 0.1568627450980392}
\definecolor{color5}{rgb}{0.5803921568627451, 0.403921568627451, 0.7411764705882353}
\definecolor{color6}{rgb}{0.5490196078431373, 0.33725490196078434, 0.29411764705882354}
\definecolor{color7}{rgb}{0.8901960784313725, 0.4666666666666667, 0.7607843137254902}
\definecolor{color8}{rgb}{0.7372549019607844, 0.7411764705882353, 0.13333333333333333}
\definecolor{color9}{rgb}{0.09019607843137255, 0.7450980392156863, 0.8117647058823529}

\newcommand{\xWalberla}{waLBerla\xspace}
\newcommand{\walberla}{{\mdseries \scshape \xWalberla}\xspace}
\newcommand{\mesapd}{\mbox{MESA-PD}\xspace}

\jyear{2022}

\begin{document}

\title{Effect of Sediment Form and Form Distribution on Porosity: A Simulation Study Based on the Discrete Element Method}


\author*[1,2]{\fnm{Christoph} \sur{Rettinger}}\email{christoph.rettinger@fau.de}
\author[1,3]{\fnm{Ulrich} \sur{R\"ude}}\nomail
\author[2]{\fnm{Stefan} \sur{Vollmer}}\nomail
\author[4]{\fnm{Roy M.} \sur{Frings}}\nomail

\affil[1]{\orgdiv{Chair for System Simulation}, \orgname{Friedrich-Alexander-Universit\"at Erlangen-N\"urnberg}, \orgaddress{\street{ Cauerstra\ss e 11}, \city{Erlangen}, \postcode{91052}, \country{Germany}}}

\affil[2]{\orgdiv{Department of Fluvial Morphology, Sediment Dynamics and Management}, \orgname{Federal Institute of Hydrology}, \orgaddress{\street{Am Mainzer Tor 1}, \city{Koblenz}, \postcode{56068}, \country{Germany}}}

\affil[3]{\orgname{CERFACS}, \orgaddress{\street{42 Avenue Gaspard Coriolis}, \city{Toulouse Cedex 1}, \postcode{31057}, \country{France}}}

\affil[4]{\orgdiv{Rijkswaterstaat Zuid-Nederland}, \orgname{Ministry of Infrastructure and Water Management}, \orgaddress{\street{Avenue Ceramique 125}, \city{Maastricht}, \postcode{6221 KV}, \country{Netherlands}}}

\abstract{
	Porosity is one of the key properties of dense particle packings like sediment deposits and is influenced by a multitude of grain characteristics such as their size distribution and shape.
	In the present work, we focus on the form, a specific aspect of the overall shape, of sedimentary grains in order to investigate and quantify its effect on porosity, ultimately deriving novel porosity-prediction models.
	To this end, we develop a robust and accurate simulation tool based on the discrete element method which we validate against laboratory experiments.
	Utilizing digital representations of actual sediment from the Rhine river, we first study packings that are composed of particles with a single form.
	There, the porosity is found to be mainly determined by the inverse equancy, i.e., the ratio of the longest to the smallest form-defining axis.
	Only for small ratios, additional shape-related properties become relevant, as revealed by a direct comparison to packings of form-equivalent ellipsoids.
	Since sediment naturally features form mixtures, we extend our simulation tool to study sediment packings with normally-distributed forms.
	In agreement with our single form studies, the porosity depends primarily on the inverse of the mean equancy.
	By supplying additional information about a second form factor and the standard deviations, we derive an accurate model for porosity prediction.
	Due to its simplicity, it can be readily applied to sediment packings for which some measurements of flatness and elongation, the two most common form factors, are available.
	}

\keywords{porosity, particle packing, shape, discrete element method, fluvial sediment}

\maketitle

\section{Introduction}

The porosity of packed particulate material, defined as the ratio of pore volume to total volume, is of interest in many natural processes or technical applications.
Taking fluvial sediment as an illustrative example, porosity relates bed level changes to the amount of transported material via the Exner equation \cite{coleman2009} and characterizes the quality of reproduction habitats for fish \cite{noack2017}.
Consequently, river engineers require accurate information about porosity and its spatial variability in those systems.
Assessing porosity directly via field studies, however, is often cumbersome and associated with large costs and thus not practicable as a general measure \cite{seitz2018,tabesh2019,frings2011,tabesh2022}.
Instead, extracting the major influence factors of porosity and combining them in an accurate and concise predictive model is a promising alternative and constitutes our main motivation of the present work. 
A multitude of parameters, ranging from individual particle properties to features of the deposition process, are known to affect porosity or, likewise, the packing fraction \cite{latham2002,fraser1935}.

Among those, the effect of particle size distribution has received the most attention, not only in engineering disciplines but also generally in mathematics and physics. 
One reason for that is that the size of a particle is often immediately and intuitively accessible and well-defined.
In the fluvial context, sieving a sample through a hierarchy of sieves is the standard method to quantify the size distribution, reported as the histogram-like sieving curve. 
These studies led to the development of mathematical models for specific size distributions \cite{brouwers2014}, empirical models based on certain characteristics like the median grain diameter $D_{50}$ or the standard deviation \cite{carling1982,wu2006,wooster2008,frings2011}, or more general theoretical mixing models \cite{yu1991,nunezGonzales2016}.
Still, when applying these porosity models for sediment deposits and comparing their prediction to experimental measurements, their predictive capabilities are rather limited \cite{seitz2018,frings2011}, suggesting that other effects interfere in the general case.

One of these is the actual shape of the packed particles.
It, too, varies on a grain by grain basis for actual sediment which poses an additional challenge, since the assessment has to happen for a large number of individual grains, making it a laborious task in field studies.
Additionally, the shape encompasses several grain properties that can be measured and classified differently \cite{bagheri2015,stuckrath2006}.
In the context of sedimentology, due to the huge variety of natural rocks, various form factors and shape parameters have been proposed and are used for classification \cite{blott2008,illenberger1991,oakey2005}.
There, the form describes the overall appearance in categories like flat or elongated \cite{zingg1935} while, in addition to it, shape also considers surface-localized characteristics like roundness and sphericity, as reviewed extensively by \cite{blott2008}. 
Like size, the form and shape of fluvial sediment may vary significantly between different rivers or spatial location inside a single river due to changes in the lithology, transportation history, and deposition environments. 

In contrast to size, however, quantifying the effect of form and shape on porosity has received significantly less attention. 
For idealized geometries like ellipsoids or superellipsoids, it has been found that the aspect ratio, the ratio of smallest and largest semi-axis, alters the porosity \cite{donev2004,delaney2010}.
These studies typically feature packings of a single shape only, and recently some advances were made by considering discrete mixtures of a few distinct shapes \cite{li2020,yuan2020}.
Regarding the shape of natural rocks, increasing roundness was observed to reduce porosity \cite{cho2006,altuhafi2016} and has thus been incorporated in predictive models \cite{peronius1985,chang2018,latham2002}, while the role of sphericity is less clear \cite{cho2006,vepraskas1987}.
Recently, \cite{zhao2020} reported that elongation and flatness, both common descriptors of the sediment form, affect the porosity of rock aggregate deposits whereas \cite{suhr2020} observed no significant correlation between these form factors and porosity for railway ballast.
Overall, the general effect of form and other shape-related properties has not yet been systematically studied for natural deposits. 
It remains unclear which of the several form factors are actually relevant for porosity and to what extent additional shape-related factors need to be considered.
Such insight, however, might guide the measurement strategies of field studies and improve predictive models.
Additionally, like size, form and shape naturally exhibit a certain distribution \cite{oakey2005,suhr2020,zhao2021}.
How such a continuous variation, in contrast to packings with a single shape, alters the porosity is also mostly to be studied and can give answers on how it could be quantified and included in a model.
An obvious reason for the absence of such shape-focused studies is that experiments with actual rocks face the formidable challenge of obtaining several thousand grains with roughly the same size and shape properties \cite{fraser1935}.
As a consequence of this inaccessibility, the effects of size and shape on porosity cannot be treated independently and are thus usually combined in the studies and models \cite{peronius1985,vepraskas1987}.

Numerical simulations that aim to provide a virtual replacement of the physical processes during packing offer several compelling advantages in this respect \cite{latham2002,coetzee2016}.
Here, a single sediment grain can infinitely be replicated, restricting the analysis to a unique but complex shape \cite{zhao2020}, or distributions of shapes and forms can be generated \cite{wang2021}. 
Simulative studies also allow to tackle the porosity-increasing effect of bounding walls \cite{zou1995,liu2000}, as present in many laboratory studies \cite{frings2011}, by using periodicity arguments and by advanced evaluation procedures since detailed information about particle position is available.
Among the various available simulation approaches, physics-based methods like the discrete element method (DEM) are able to account for the desposition process and frictional interactions \cite{lu2015,seelen2018}.
They are thus becoming increasingly popular to study dense particulate systems and their macroscopic quantities like porosity and shear behavior \cite{zhao2017,gong2019,zhao2020,tong2015,bui2019}.
Despite challenges like the treatment of non-spherical particles \cite{seelen2018} combined with the computational effort required to simulate a large enough number of particles \cite{preclik2015}, the striking advantage of such simulation-based studies is a direct control over all aspects of the packing process.
The compute power offered by current high-performance computing (HPC) machines enables realistic packing scenarios and parametric studies.

We thus aim to develop an accurate simulation tool based on actual sediment shapes to study dense packings and evaluate their porosity.
After validation against experimental results, this tool is used for systematic studies of the effect of grain size and form, where both properties can be controlled and quantified individually.
These results are then combined to explore and quantify relationships between grain properties and porosity, yielding porosity prediction models for single form and form distribution packings.
Due to the wide range of natural rock sizes and shapes, we primarily focus on fluvial gravel sediment, noting that most findings can readily be applied for other fields.

This paper is thus structured as follows. 
First, we introduce the form classification in \cref{sec:sedimentary_grains} and carry out an in-depth analysis of digitized sediment, assessing their form distribution.
The next three sections focus on establishing a robust and validated tool for numerical packing simulations. 
In \cref{sec:method}, we present the numerical method of our simulation tool that makes use of the DEM to account for realistic particle interactions. 
We then describe the two distinct simulation domains used in our studies, the simulation parameterization and process to generate dense packings, and the porosity evaluation routine in \cref{sec:simulation_processes}.
\Cref{sec:validation} calibrates and validates the simulation approach via comparisons to laboratory experiments in a cylindrical domain.
The then following two sections apply this tool and feature the main studies regarding the influence of grain form on porosity. 
In \cref{sec:study_single}, we simulate packings with sediment and ellipsoidal particles of a single form and correlate the obtained porosity to the form factors to derive a predictive model.
These studies are extended to packings with form distributions in \cref{sec:study_mixture}, for which again a predictive model is derived.
\Cref{sec:conclusion} summarizes the main findings and outlines future directions.

\section{Size and Form Parameters of Sedimentary Grains}

\label{sec:sedimentary_grains}

In this section, we introduce the three major parameters that are used to define size and form, stated as various form factors, of a single grain.
We investigate the form distribution and form factors for a set of digitized fluvial sediment grains.

\subsection{Grain Description}


In the context of sedimentology, a single grain is commonly described by three orthogonal measures.
They are denoted as the long $L$, intermediate $I$, and short $S$ dimension, implying that $L\geq I \geq S$.
As reviewed in \cite{blott2008}, different approaches for their determination exist and, consequently, the results may vary.
Often, one tries to find the smallest box that fully encloses the grain whose side lengths then denote the three form parameters \cite{zingg1935}.
A modification was proposed by \cite{stuckrath2006}, where the three axes intersect at the grain's center of mass and they are defined by the major axes of an ellipsoid, whose mass and moments of inertia are identical to the grain's.
The latter makes the determination insensitive to irregular surface contours, focusing solely on the form.
We thus follow this procedure here.



\begin{table*}
	\centering
	\begin{tabular}{llll}
		\toprule
		Form factor & Abbreviation & Definition & Range\\\midrule
		flatness & $\mathit{FL}$ & $S/I$ & $(0,1]$\\
		elongation & $\mathit{EL}$ & $I/L$ & $(0,1]$\\
		equancy & $\mathit{EQ}$ & $S/L$ & $(0,1]$ \\ 
		Aschenbrenner shape factor & $\mathit{ASF}$ & $L\,S / I^2$ & $(0,\infty)$ \\
		Corey shape factor & $\mathit{CSF}$ &  $S / \sqrt{L\,I}$ & $(0,1]$\\
		disc-rod index & $\mathit{DRI}$ & $(L-I)/(L-S)$ & $(0,1)$\\
		Illenberger form index & $\mathit{IFI}$ & $(L+S) / I $ & $(1,\infty)$\\ 
		Illenberger rod index & $\mathit{IRI}$ & $(S+I) / L $ & $(0,2]$\\
		Janke form factor & $\mathit{JFF}$ & $S/\sqrt{(L^2+I^2+S^2)/3}$ & $(0,1]$\\
		Krumbein intercept sphericity & $\mathit{KIS}$ & $\sqrt[3]{(I\,S)/L^2}$ & $(0,1]$\\
		maximum projection sphericity & $\mathit{MPS}$ & $\sqrt[3]{S^2/(L\,I)}$ & $(0,1]$\\
		oblate-prolate index & $\mathit{OPI}$ & $10((L-I)/(L-S) - \tfrac{1}{2})/(S/L)$ & $(-\infty,\infty)$\\
		Wentworth flatness index & $\mathit{WFI}$ & $(L+I) / (2S) $ & $[1,\infty)$\\
		\bottomrule
	\end{tabular}
	\caption{Form factors commonly used in sedimentological fields \cite{illenberger1991,blott2008}.} 
	\label{tab:form_factors}
\end{table*}

It is important to clearly separate grain size and form since both pieces of information are contained in $L$, $I$ and $S$.
The grain size of a sample is typically determined by a hierarchy of sieves with gradually finer meshes.
This procedure poses geometric constraints on the grains that pass through a certain sieve mesh.
For that reason, we use a sieve-based size definition where the grain size $D$ is given by \cite{oakey2005}
\begin{equation}
	D = \sqrt{\frac{S^2+I^2}{2}}, \label{eq:grain_size}
\end{equation}
corresponding to the edge width of the bounding square of an ellipse with height $S$ and width $I$.
A useful quantification of grain form, on the other hand, should be independent of the actual size, implying that it is dimensionless.
These so-called \emph{form factors} can be constructed by combining $L$, $I$ and $S$ and always contain ratios of these parameters.
In the literature, a huge variety of these form factors can be found, as e.g. reviewed in \cite{illenberger1991,blott2008}.
We summarize the most important ones in \cref{tab:form_factors}.
Mathematically, only two of them are strictly required to uniquely define the form of a single grain~\cite{illenberger1991}.
Based on the work of \cite{zingg1935}, a common choice is the pair of flatness and elongation, leading to the so-called Zingg diagram.


\subsection{Form of Fluvial Sediment}
\label{sec:form_meshes}


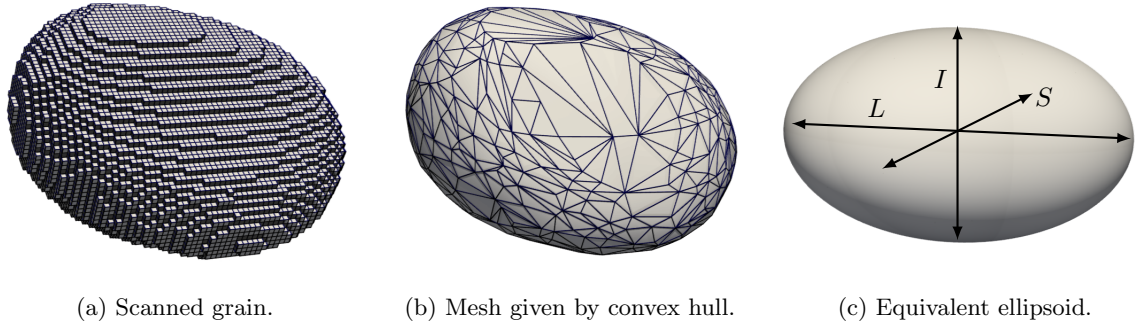
\begin{figure*}
	\begin{subfigure}{0.32\textwidth}
	\includegraphics[trim=450 300 450 300,clip,width=\textwidth]{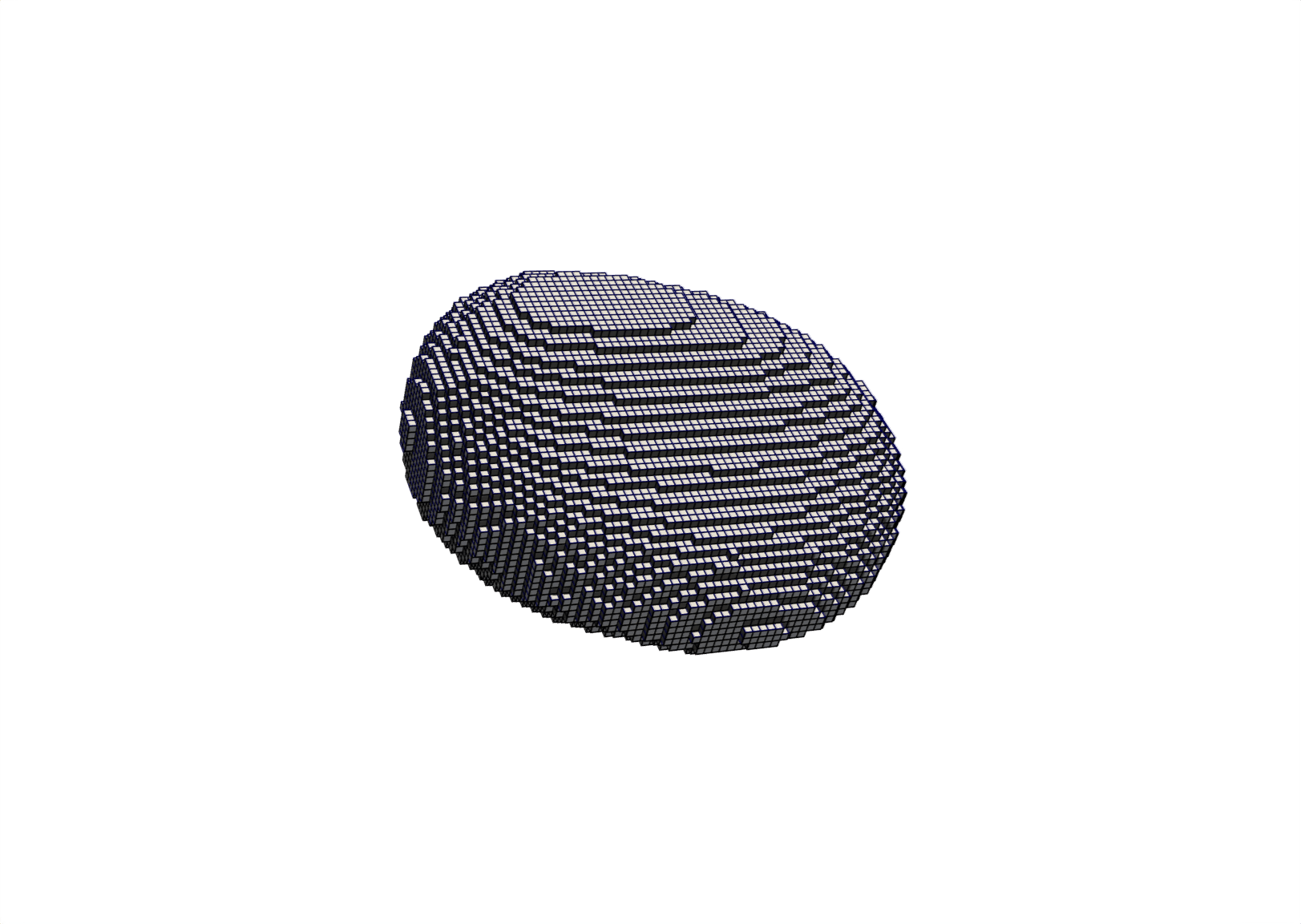}
	\caption{Scanned grain.}
	\end{subfigure}	
\begin{subfigure}{0.32\textwidth}
	\includegraphics[trim=450 300 450 300,clip,width=\textwidth]{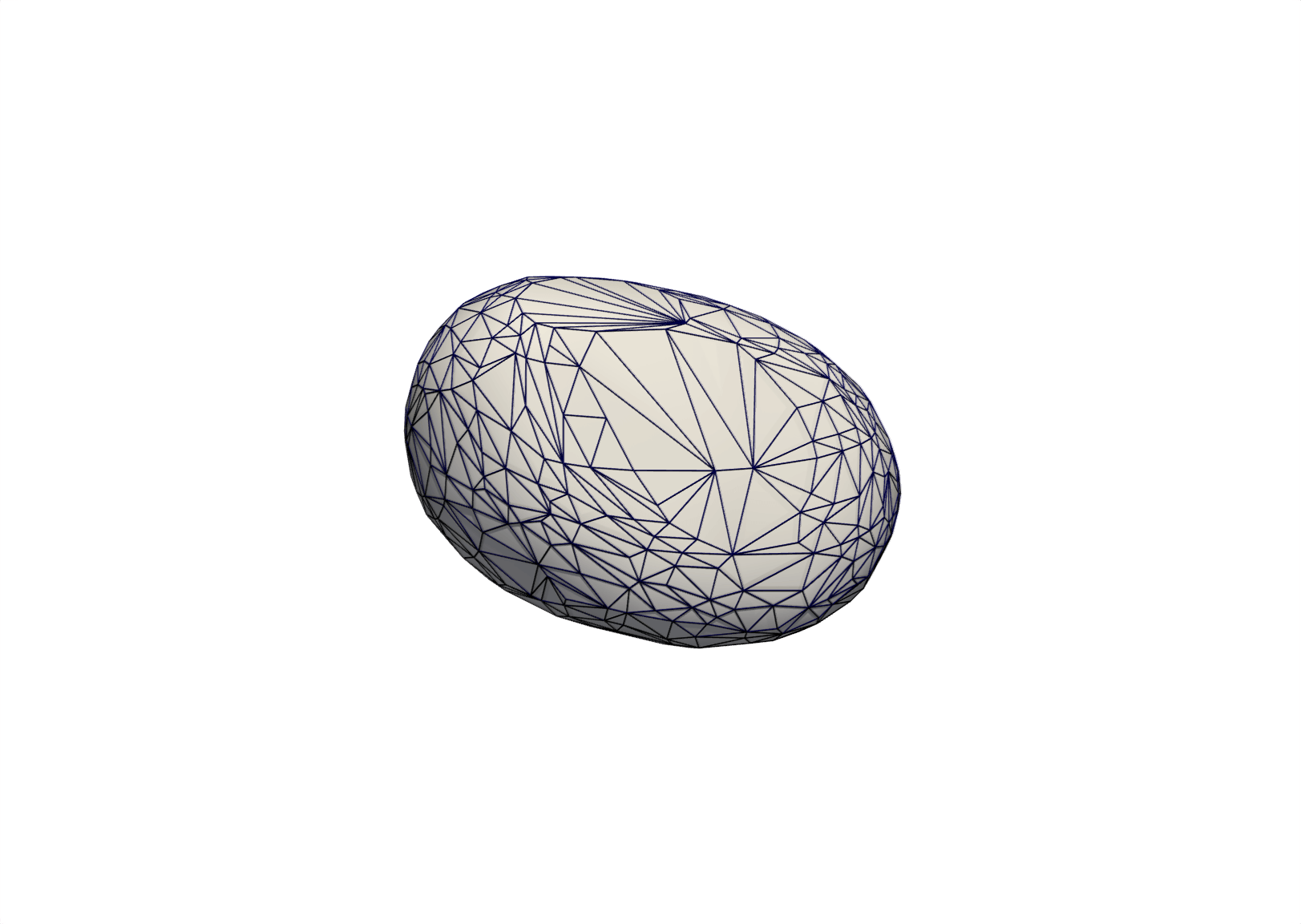}
	\caption{Mesh given by convex hull.}
\end{subfigure}
	\begin{subfigure}{0.32\textwidth}
	\begin{tikzpicture}
	\node[inner sep=0pt] (vis) at (0,0) {\includegraphics[trim=450 300 450 300,clip,width=\textwidth]{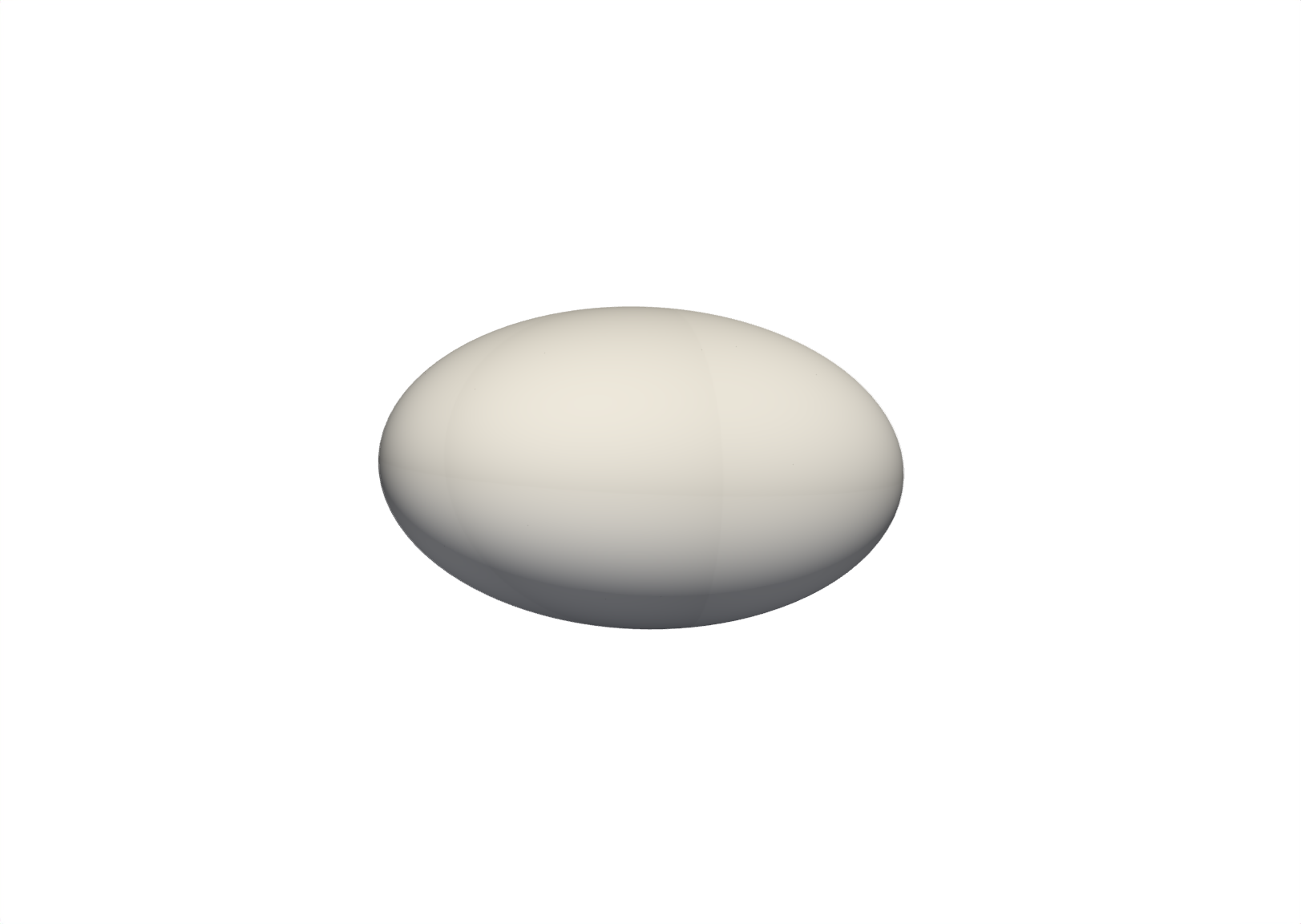}};
	\draw[thick, latex-latex] (-2.3,0.1) -- (2.2,-0.1) node[pos=0.25,above,color=black, align=center]{$L$};
	\draw[thick, latex-latex] (-0.1,1.4) -- (-0.1,-1.45) node[pos=0.25,left,color=black, align=center]{$I$};
	\draw[thick, latex-latex] (-1.1,-0.5) -- (0.9,0.5) node[pos=0.95,right,color=black, align=center]{$S$};
	\end{tikzpicture}
	\caption{Equivalent ellipsoid.}
\end{subfigure}
\caption{Data processing for the generation of the surface mesh and the evaluation of the form parameters.}
\label{fig:mesh_processing}
\end{figure*}

In this work, we made use of 63 digitized grains that were obtained via computer tomography scanning \cite{liang2015}.
The scanned samples are fluvial gravels collected from the Rhine River in Germany and randomly selected from seven different size fractions, i.e., 2.8-4 mm, 4-5.6 mm, 5.6-8 mm, 8-11.2 mm, 11.2-16 mm, 16-22.4 mm, and 22.4-31.5 mm.
From such a scan, we obtained the subset of the equally spaced three dimensional sample space of points that are contained inside the grain.
We then constructed the convex hull of this point cloud as a robust measure to determine a triangulated surface mesh.
With the help of the Python library trimesh \cite{trimesh}, we moved the mesh's center of mass into the origin of the coordinate system, rotated it such that the principal axes are aligned with the coordinate axes, and computed the moments of inertia.
As mentioned before, these moments determine the equivalent ellipsoid and, consequently, $L$, $I$, and $S$.
These steps are shown in \cref{fig:mesh_processing} and we provide the details about this procedure in \cref{sec:app:mass_equivalent_ellipsoid}.

\begin{figure*}
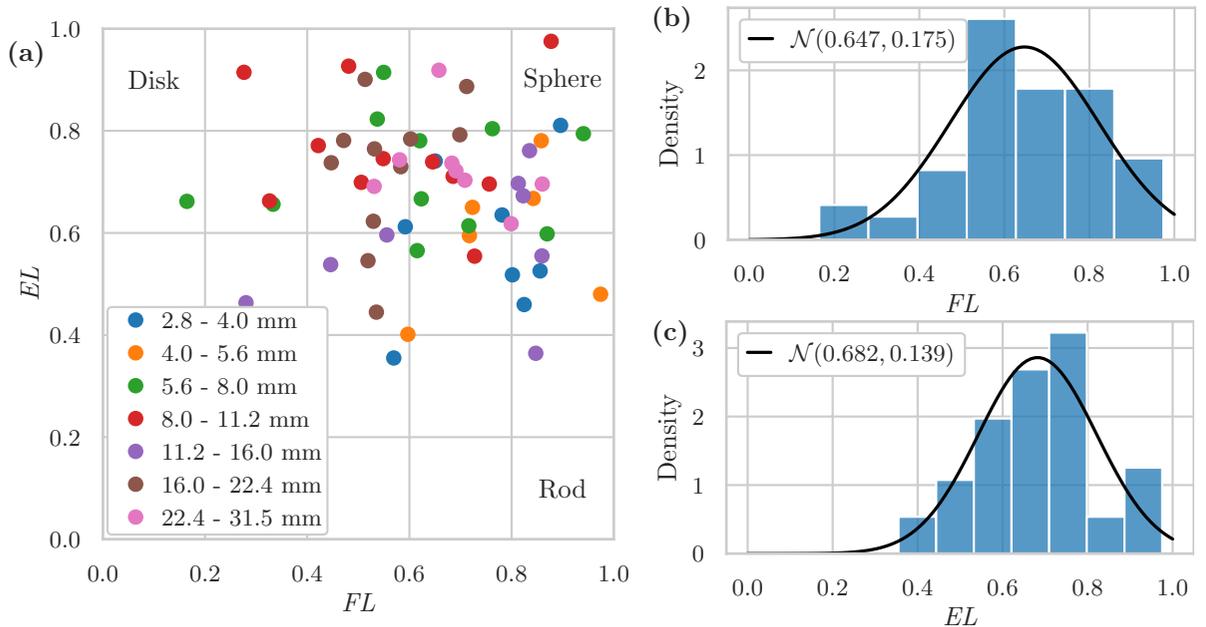

	\centering
	\begin{subfigure}{0.54\textwidth}
	\input{figures/grainsElongationFlatness.pgf}
	\end{subfigure}~
	\begin{subfigure}{0.45\textwidth}
	\begin{subfigure}{\textwidth}
			\input{figures/flatnessDistribution.pgf}
		\end{subfigure}
		\begin{subfigure}{\textwidth}
			\input{figures/elongationDistribution.pgf}
		\end{subfigure}
	\end{subfigure}
	\caption{Flatness (\textit{FL}) and elongation (\textit{EL}) of the scanned Rhine sediment. \textbf{a} Zingg diagram where the color encodes the size class. Distributions, given as histogram and maximum-likelihood estimation of an assumed Gaussian distribution $\mathcal{N}(\mu,\sigma)$, of \textbf{b} flatness and \textbf{c} elongation.}
	\label{fig:mesh_flatness_elongation}
\end{figure*}


Next, we evaluated the size, flatness, and elongation of all 63 available grains in this way.
This analysis is shown in \cref{fig:mesh_flatness_elongation} on the left, distinguishing the different size fractions.
The majority of the data points is found in the upper right region of the diagram, which is classified as sphere-like, while some also exhibit a disc-like form.
Evaluating the correlation coefficient according to Pearson for the size and the two form factors, we obtained $r=-0.11$ for flatness and $r=0.23$ for elongation, indicating no significant dependence.
This finding agrees well with the visual inspection of \cref{fig:mesh_flatness_elongation}.
Consequently, we assumed that size and form are independent quantities in our studies.


Histograms of the distribution of flatness and elongation for all grains are shown in the right panels of \cref{fig:mesh_flatness_elongation}.
Both distributions exhibit a normal distribution, according to the Shapiro-Wilk test with a p-value of 0.05. 
The corresponding mean and standard deviation values are given in the legend.
This observation is in line with other studies, e.g., of granite rock aggregates by \cite{zhao2020}.
Additionally, we gathered and visually extracted sets of flatness - elongation data from the literature, checked them for normality and extracted values of mean and standard deviation.
The findings of this study are summarized in \cref{tab:shape:flatness_elongation_literature}.
Note that especially the manually measured samples might exhibit certain biases due to the operator \cite{blott2008}, together with the general sampling bias and the applied form quantification technique.
Despite the different geological origins of the distinct samples, we observe normally distributed flatness and elongation values in most cases.
The mean and standard deviation of flatness are generally in the ranges of $[0.35, 0.8]$ and $[0.13, 0.2]$, respectively, whereas the elongation values are found in narrower ranges of $[0.68,0.77]$ and $[0.12, 0.17]$.


\begin{table*}[t]
	\centering
	\begin{tabular}{llllllll}
		\toprule
		Data & Type & Size range (mm) & $N_s$ & $\mu_\mathit{FL}$ & $\sigma_\mathit{FL}$ & $\mu_\mathit{EL}$ & $\sigma_\mathit{EL}$  \\\midrule
		\cref{fig:mesh_flatness_elongation} & river (Rhine) & 2.8 - 31.5 & 63 & 0.647 & 0.175 & 0.682 & 0.139 \\
		\cite{liang2015} & river (Kall) & 2.8 - 31.5 & 63 & 0.357(*) & 0.131(*) & 0.710 & 0.168 \\
		\cite{oakey2005} & river (Lune) & 8 - 182  & 142 & 0.555 & 0.178 & 0.728 & 0.150 \\ 
		\cite{zhao2020} & rock aggregate & 31.5 - 90 & 1000 & 0.790 & 0.207 & 0.714 & 0.153 \\ 
		\cite{blott2008} & gravel & 5.6 - 8 & 93 & 0.660 & 0.158 & 0.746 & 0.134 \\ 
		 & gravel & $>$ 8 & 94 & 0.633 & 0.175 & 0.761(*) & 0.121(*) \\
		\cite{suhr2020} & railway ballast & 20 - 40 & 50 & 0.707(*) & 0.191(*) & 0.707 & 0.132 \\ 
		\cite{barrett1980} & glacier & 8 - 64 & 597 & 0.59 & 0.15 & 0.72 & 0.13 \\ 
		 & glacier & 8 - 64 & 706 & 0.67 & 0.15 & 0.75 & 0.12 \\ 
		\bottomrule
	\end{tabular}
\caption{Distribution, given as mean and standard deviation, of flatness and elongation reported in literature. Values marked with (*) indicate cases where the hypothesis of normal distribution has been rejected by the test. $N_s$ is the number of samples in each data set.}
\label{tab:shape:flatness_elongation_literature}
\end{table*}

\subsection{Correlation of Form Factors}
\label{sec:shape:correlation}

\begin{figure}
	\centering
	\input{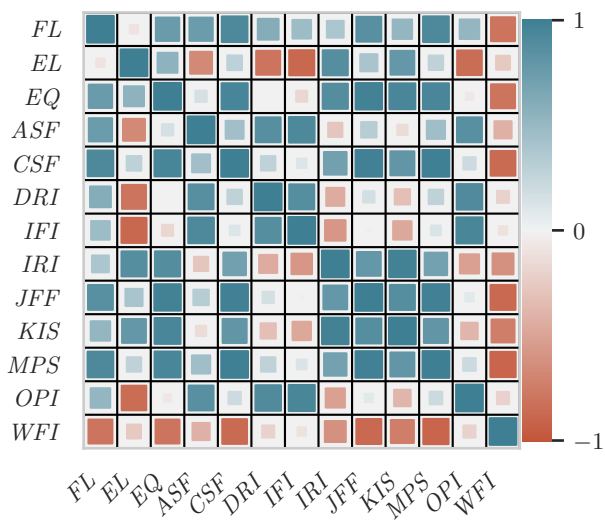}
	\caption{Correlation coefficients of form factors from \cref{tab:form_factors} for the scanned Rhine sediment.}
	\label{fig:formFactorCorrelation}
\end{figure}

Despite the multitude of available form factors, they essentially describe similar properties as they all depend on just three parameters $L,I,$ and $S$.
As such, they are not independent features but might exhibit strong correlations, as e.g. analyzed for volcanic particles~\cite{bagheri2015} and for railway ballast~\cite{suhr2020}.
To illustrate this aspect for our shapes, important for further data analytical investigations and models, we evaluated the form factors stated in \cref{tab:form_factors} for the 63 available grain geometries and computed their correlation according to Pearson.
This analysis is shown in \cref{fig:formFactorCorrelation} as a correlation matrix.
While there are pairs with low correlations, like flatness and elongation with a value of $r=-0.089$, a large portion of all entries shows absolute values close to 1, indicating a strong correlation.
We also conducted a principle component analysis of the form factors and found that only two of them are required to explain the variance in these features.
This is in line with our expectation based on mathematical reasoning about the number of necessary form factors.

\section{Numerical Method for Particle Simulation}
\label{sec:method}

In this section, we present the numerical method that we applied to study dense packings of sedimentary particles via simulations. 
We made use of the \mesapd framework \cite{eibl2021thesis}, integrated into the \walberla multi physics framework \cite{walberla2020}.
This open-source software offers functionality for accurate particle simulations, provides mechanisms that allow for flexible extensions of the models, and enables large-scale execution on HPC supercomputers.
In particular, the implementation of the numerical method as applied here and the simulation setup of our studies can be found in the main repository\footnote{\url{https://walberla.net}}.

\subsection{Particle Motion and Collision Model}

For a physical representation of the packing behavior, we used the particle model that has already been successfully applied for coupled fluid-particle simulations of sediment beds \cite{rettinger2022}.
A rigid particle is described by its mass $m_p$, moment of inertia tensor $\mathbf{I}_{p}$, 
center of mass  $\boldsymbol{x}_{p}$, rotation $\boldsymbol{\theta}_{p}$ (here represented by quaternions), and translational and rotational velocities, $\boldsymbol{u}_{p}$ and  $\boldsymbol{\omega}_{p}$.
The motion of a particle $i$ is governed by the Newton-Euler equations of motion, generally given as
\begin{align}
\frac{\text{d} \boldsymbol{x}_{p,i}}{\text{d}t} = \boldsymbol{u}_{p,i},\quad m_{p,i}  \frac{\text{d} \boldsymbol{u}_{p,i}}{\text{d}t} =  \boldsymbol{F}_{p,i}^\mathit{col} + \boldsymbol{F}_{p,i}^\mathit{ext}, \label{eq:method:particle_translation}\\
\frac{\text{d} \boldsymbol{\theta}_{p,i}}{\text{d}t} = \boldsymbol{\omega}_{p,i},\quad \frac{\text{d}(\mathbf{I}_{p,i} \boldsymbol{\omega}_{p,i})}{\text{d}t} = \boldsymbol{T}_{p,i}^\mathit{col}, \label{eq:method:particle_rotation}
\end{align}
where the force $\boldsymbol{F}_{p,i}^\mathit{col}$ and torque $\boldsymbol{T}_{p,i}^\mathit{col}$ account for collisions with other particles or walls, and $\boldsymbol{F}_{p,i}^\mathit{ext}$ combines external forces, e.g., due to gravity.
We used a semi-implicit Euler scheme to integrate \cref{eq:method:particle_translation,eq:method:particle_rotation} in time with a time step size of $\Delta t$ that first updates the velocities and then uses these values to compute the new position and rotation \cite{wachs2019}.

\begin{figure}[t]
	\centering
	\begin{tikzpicture}[scale=0.8]
	\coordinate[label=left:$\boldsymbol{x}_{p,i}$] (xpi) at (2,2);
	\coordinate[label=right:$\boldsymbol{x}_{p,j}$] (xpj) at (5.4,2.5);
	
	\coordinate[label={[label distance=5]right:$\boldsymbol{x}_{ij}^\mathit{cp}$}] (C) at ($ (xpj)!0.5!(xpi) $ );
	\coordinate (nij) at ($0.5*(xpi)-0.5*(C)$ );
	\coordinate (tij) at ([rotate=90]nij);
	\draw[brown,thick,fill=brown, fill opacity=0.2] ($(xpi)-2.3*(nij)$) -- (3.5,3.5) -- (0.5,3.4) -- (0,1.7) -- (2,0.5) -- (3,0.6) -- cycle;
	\draw[brown,thick,fill=brown, fill opacity=0.2] ($(xpj)+2.3*(nij)$) -- (3.9,4) -- (5.8,4.5) -- (7.5,2.3) -- (6,0.8) -- (3.6,1) -- cycle;
	
	\node[fill=black, circle, inner sep=1.5] at (xpi) {};
	\node[fill=black, circle, inner sep=1.5] at (xpj) {};
	\node[fill=black, circle, inner sep=1.5] at (C) {};
	
	\draw[-latex, thick] (C) -- ++(nij) node[pos=1,below]{$\boldsymbol{n}_{ij}$};
	\draw[-latex, thick] (C) -- ++(tij) node[pos=1,right]{$\boldsymbol{t}_{ij}$};
	\draw[-latex, thick] (xpi) -- ++(0.5,-0.9) node[pos=1,right]{$\boldsymbol{u}_{p,i}$};
	\draw[-latex, thick] (xpj) -- ++(-0.4,1.1) node[pos=1,left]{$\boldsymbol{u}_{p,j}$};
	
	\coordinate (intersect1) at ($(C)-0.33*(nij)$);
	\coordinate (intersect2) at ($(C)+0.33*(nij)$); 
	\draw[blue] (intersect1) -- ++($-2.2*(tij)$);
	\draw[blue] (intersect2) -- ++($-2.2*(tij)$);
	\draw[thick,latex-latex,blue] ($(intersect1)-2*(tij)$) -- ($(intersect2)-2*(tij)$) node[pos=0.5,above]{\color{black}{$\delta_{ij,n}$}};
	
	\draw[-latex,thick] ($(xpi) + (190:2.3)$) arc (190:140:2.3);
	\node[left] at ($(xpi) + (140:2.3)$) {$\boldsymbol{\omega}_{p,i}$};
	\draw[-latex,thick] ($(xpj) + (-10:2.3)$) arc (-10:35:2.3);
	\node[right] at ($(xpj) + (35:2.3)$) {$\boldsymbol{\omega}_{p,j}$};
	\end{tikzpicture}
	\caption{Schematic representation of two colliding particles, $i$ and $j$, together with particle and contact quantities.}
	\label{fig:method:interaction_sketch}
\end{figure}
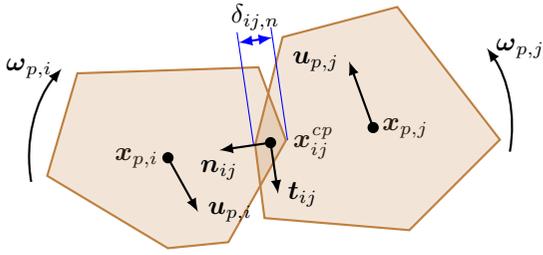

To model the collision-based interaction, we applied the discrete element method (DEM), originally proposed by \cite{cundall1979} and 
for which further details about the collision model are given in \cite{rettinger2022_jcp}.
It considers the collision response between pairs of particles and splits it into a normal and tangential part, $\boldsymbol{F}_{ij,n}^\mathit{col}$ and $\boldsymbol{F}_{ij,t}^\mathit{col}$, respectively.
The total collision force and torque on a particle $i$ are then given as
\begin{align}
\boldsymbol{F}_{p,i}^\mathit{col} = \sum_{j \neq i} \left( \boldsymbol{F}_{ij,n}^\mathit{col} + \boldsymbol{F}_{ij,t}^\mathit{col} \right),\\
\boldsymbol{T}_{p,i}^\mathit{col} = \sum_{j \neq i} (\boldsymbol{x}_{ij}^\mathit{cp}-\boldsymbol{x}_{p,i})\times \boldsymbol{F}_{ij,t}^\mathit{col}, \label{eq:DEM_total_collision}
\end{align}
where $\boldsymbol{x}_{ij}^\mathit{cp}$ is the collision point.
A sketch of two colliding particles is shown in \cref{fig:method:interaction_sketch}.

In normal direction, the force is modeled via a linear spring-dashpot model
\begin{equation}
\boldsymbol{F}_{ij,n}^\mathit{col} = -k_n \delta_{ij,n} \boldsymbol{n}_{ij} - d_n \boldsymbol{u}_{ij,n}^\mathit{cp}, 
\label{eq:method:DEM_normalCollisionForce}
\end{equation}
where $k_n$ and $d_n$ are the normal stiffness and damping coefficients, respectively.
Furthermore, $\boldsymbol{n}_{ij}$ is the contact normal, $\delta_{ij,n}$ the signed surface distance, and $\boldsymbol{u}_{ij,n}^\mathit{cp}$ the relative surface velocity at the contact point in direction of the contact normal.

Similarly, the tangential collision force is given as
\begin{equation}
\boldsymbol{F}_{ij,t}^\mathit{col} =  \min(\|- k_t \boldsymbol{\delta}_{ij,t} - d_t \boldsymbol{u}_{ij,t}^\mathit{cp}\|, \| \mu_p \boldsymbol{F}_{ij,n}^\mathit{col} \|) \boldsymbol{t}_{ij}, 
\end{equation}
where $k_t$ and $d_t$ are the tangential stiffness and damping coefficients.
This model introduces the tangential displacement $\boldsymbol{\delta}_{ij,t}$, the relative tangential surface velocity at the contact point $\boldsymbol{u}_{ij,t}^\mathit{cp}$ and the direction of the tangential force $\boldsymbol{t}_{ij}$.
The magnitude of the tangential force is limited by the Coulomb friction force which then indicates slipping of the two surfaces.
This second component is accounted for by the friction coefficient $\mu_p$.

In our simulations, we considered convex non-spherical particles that were either ellipsoids, specified by their three axes, or polygonal particles, given by a surface mesh and handled internally by the OpenMesh library~\cite{botsch2002}.
As recently reviewed by \cite{lu2015,wachs2019}, the contact detection in this case, that includes the determination of the contact point, the penetration depth, and the contact normal, is significantly more complex and computationally costly than for spherical particles.
Similar to \cite{seelen2018}, we applied a combination of the Gilbert-Johnson-Keerthi algorithm (GJK)~\cite{gjk1988,bergen1999} and the expanding polytope algorithm (EPA)~\cite{bergen2001} which has already been used successfully for large-scale granular flow simulations~\cite{preclik2015}.
This approach enabled an accurate representation of the particle surface without further calibration, in contrast to the commonly applied multi-sphere approach \cite{coetzee2016,zhao2020,suhr2020,gong2019}.
To reduce the computational effort, the contact detection was split into two phases: a cheap but approximate one, followed by a costly but accurate one.
In the first phase, the set of possible contacts was narrowed down by inexpensive checks.
Here, we made use of a linked cells data structure to only consider geometrically close particles for possible collisions. 
Then, for a particle pair, we used the enclosing spheres of both particles as a further measure to rule out non-interacting particle pairs.
Only then, we applied the costly GJK-EPA to obtain accurate contact information.

Parallel execution of the packing simulation is enabled by geometric domain partitioning in a distributed memory environment \cite{walberla2020}.
It ensures efficient simulations and scalability across a wide range of number of processes which has been demonstrated previously \cite{preclik2015,eibl2021thesis}.
For the present studies, a typical packing simulation was carried out on 100 to 400 processes for a few hours.

\subsection{Generation of Size and Shape Distributions}
\label{sec:method:generation}

\begin{figure*}[t]
	\centering
	\begin{tikzpicture}
	\node[inner sep=0pt] (vis) at (0,0) {\includegraphics[trim=150 150 150 200, clip, height=4cm]{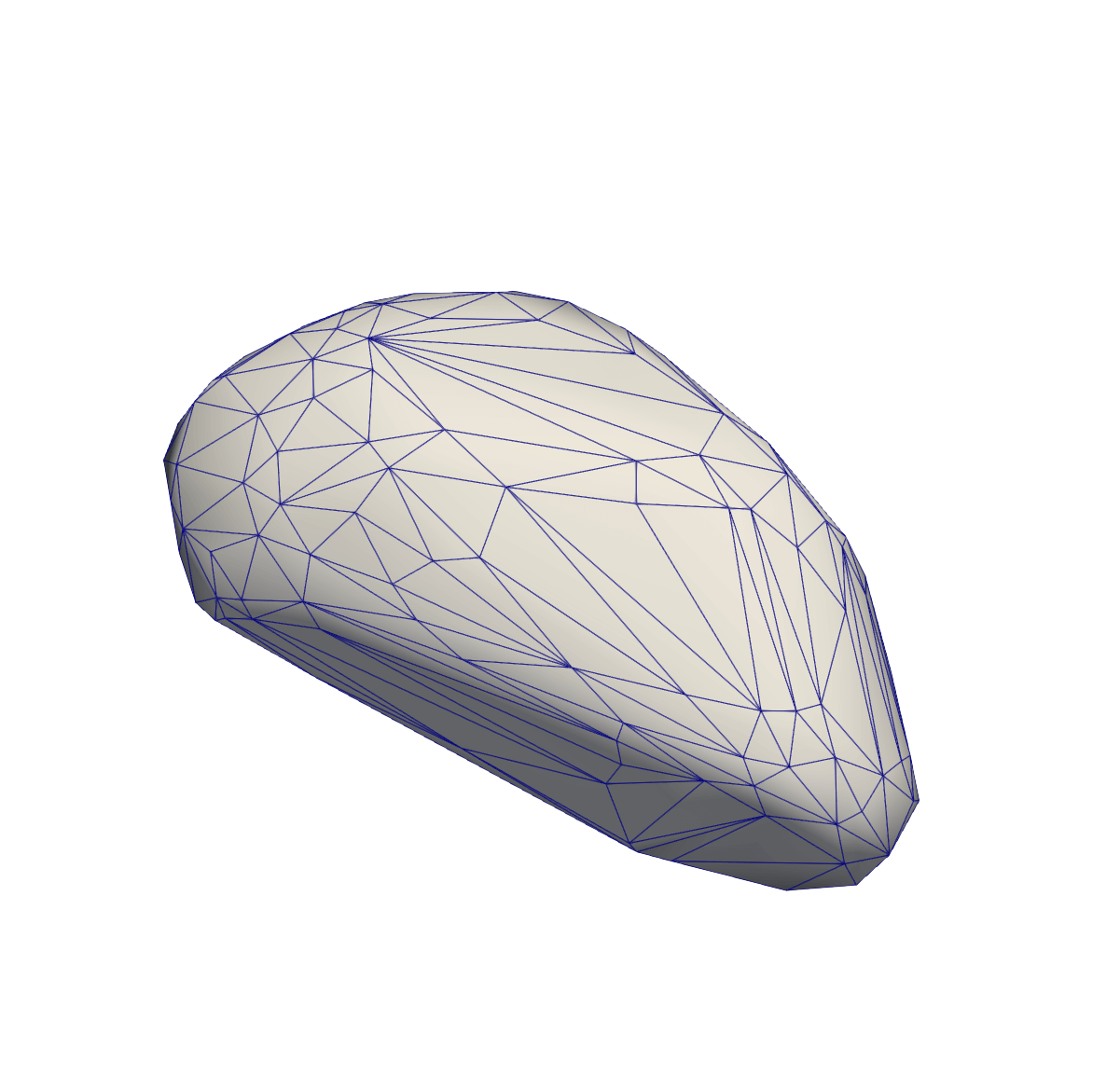}};
	\node[inner sep=0pt] (vis) at (5,0) {\includegraphics[trim=150 150 150 200, clip, height=4cm]{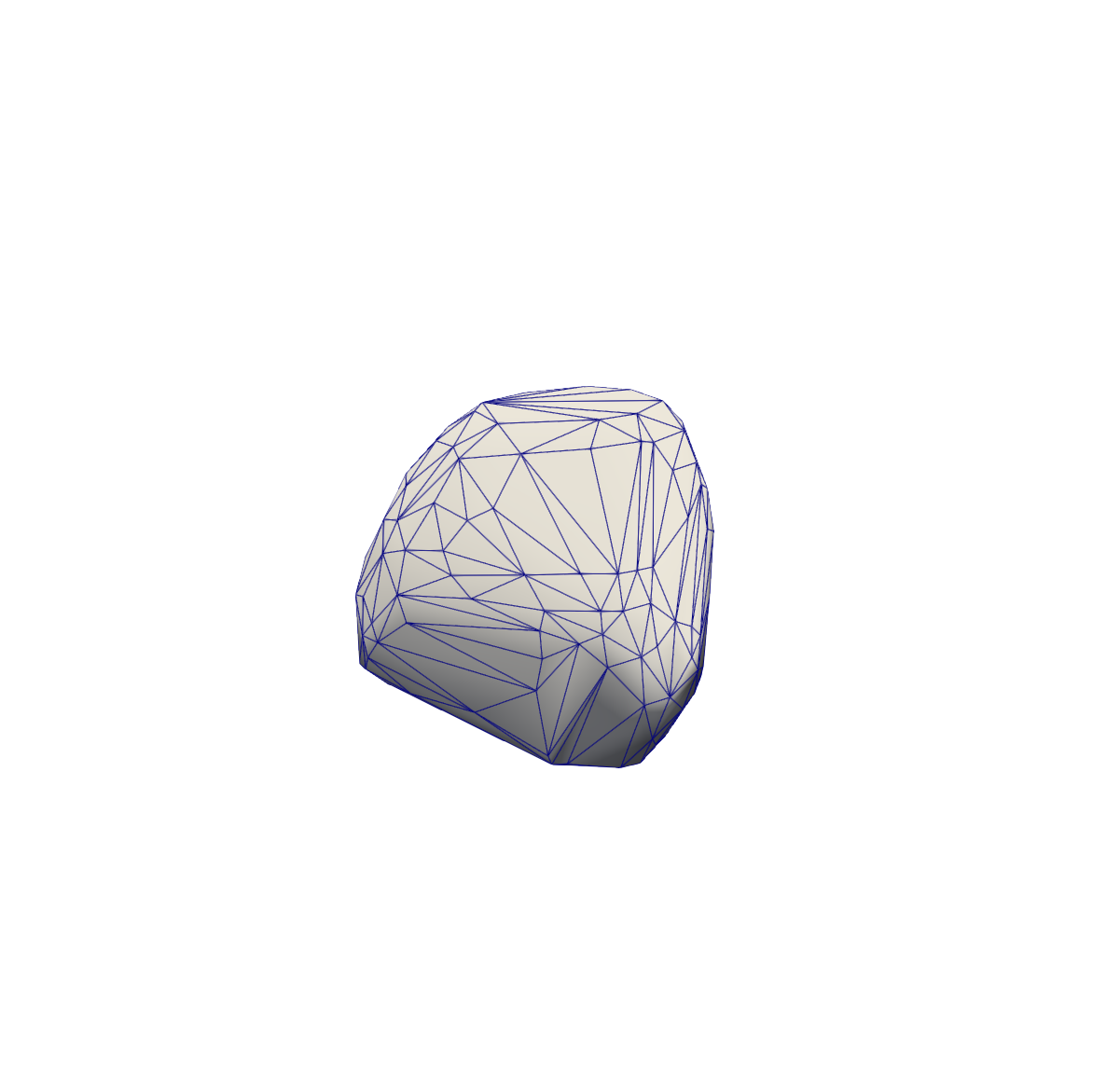}};
	\node[inner sep=0pt] (vis) at (10,0) {\includegraphics[trim=150 150 150 200, clip, height=4cm]{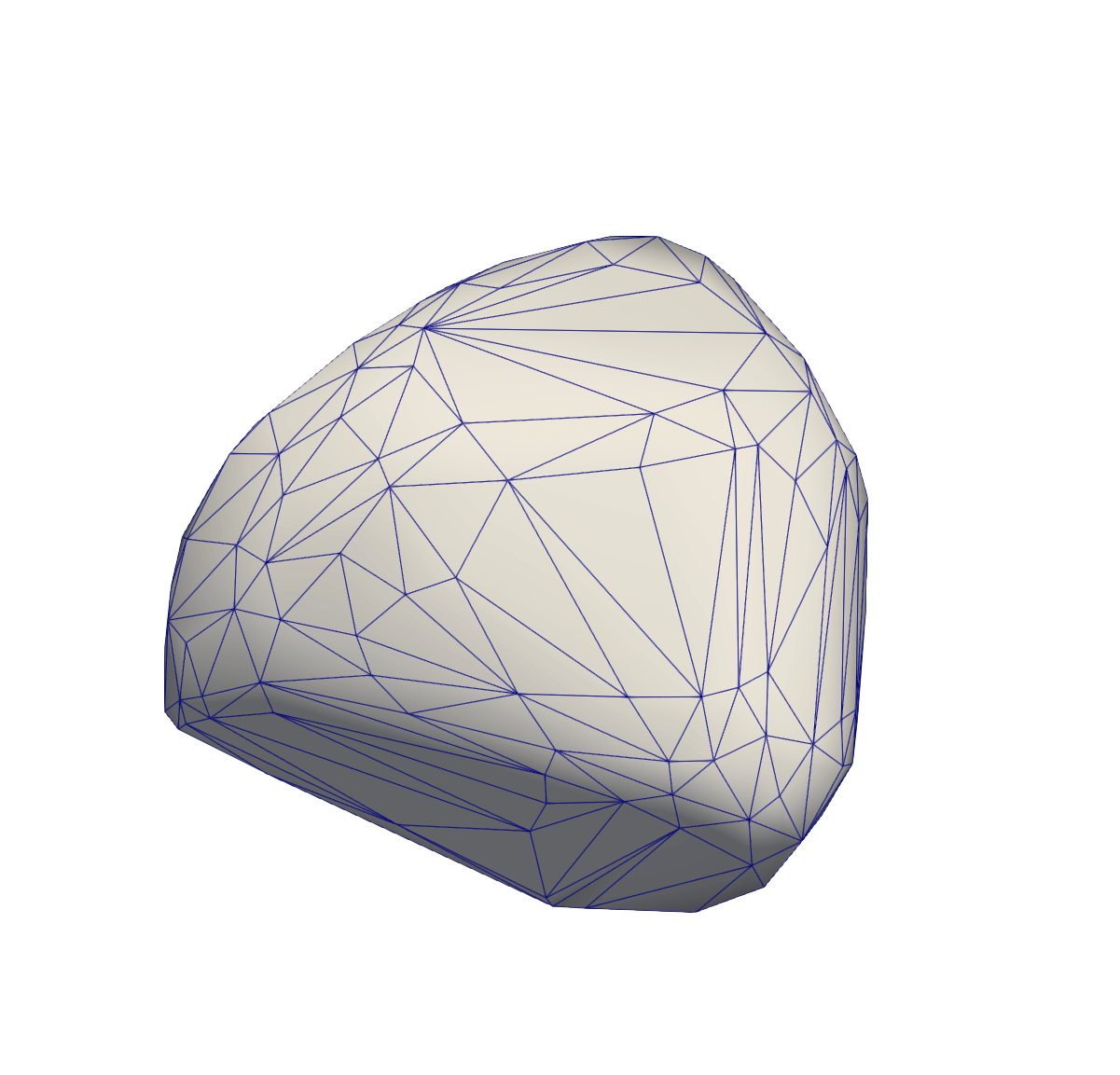}};
	\draw[thick, -latex] (1.8,0) -- ++(2,0) node[pos=0.5,above,color=black, text width=2.3cm, align=center]{scaling by $(1/L, 1/I, 1/S)$};
	\draw[thick, -latex] (6,0) -- ++(1.8,0) node[pos=0.5,above,color=black, text width=2cm, align=center]{scaling by \phantom{aa}$(\hat{L}, \hat{I}, \hat{S})$\phantom{aa}};
	\node[] at (-1.5,-1.3) {\textbf{(a)}};
	\node[] at (4,-1.3) {\textbf{(b)}};
	\node[] at (8,-1.3) {\textbf{(c)}};
	\end{tikzpicture}
	\caption{Processing of the mesh for studies with prescribed form parameters. \textbf{a} original mesh, \textbf{b} removed form factor information, \textbf{c} rescaled to match nominal form factors $\hat{L}, \hat{I}, \hat{S}$.}
	\label{fig:method:mesh_shape_scaling}
\end{figure*}
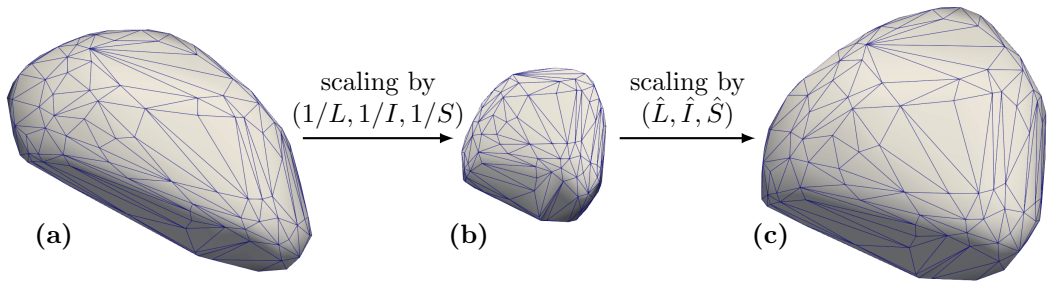

In the context of fluvial grains, information about the size is commonly given by a mass-fraction-based sieving curve.
Originating from the hierarchy of sieves, it is thus a discretized representation of the -- in reality -- continuous distribution of sizes.
Due to the large range of size values, even when considering only gravel, a base-2 logarithmic scale is typically used.
In order to convert from a mass-based to a number-based distribution, as required for numerical sampling, we used the geometric mean of each sieve class $i$:
\begin{equation}
	\bar{d}_i = \sqrt{s_i \, s_{i+1}},
\end{equation}
where $s_i$ and $s_{i+1}$ are the lower and upper sieve sizes of this size class.
The mass fractions $m_i$ can then by converted to number-equivalents $n_i$ via
\begin{equation}
n_i = m_i \, m_\mathit{tot} / ( \rho_p  \bar{V}_i),
\end{equation}
where $m_\mathit{tot}$ is the total packing mass, $\rho_p$ the grain density, and $\bar{V}_i$ is the representative grain volume per size class, which we calculated as $\bar{V}_i = \pi\bar{d}_i^3/6$.
The number-based weights then define a linear piecewise distribution on a logarithmic size scale which we applied for sampling.
We converted the thus obtained values back to actual size values which were then used to create a particle of the given size $D$.
Independent of its form, this was done by scaling the particle uniformly such that the size according to the definition from \cref{eq:grain_size} was obtained.

Regarding the form of the individual particles, we applied the assumption from \cref{sec:form_meshes} and took it to be independent of the actual size.
For single-form setups, we then selected a particular surface mesh or an equivalent ellipsoid for all particles.
For setups with grains of varying form, we usually aimed to reproduce a packing that is similar to the one from which the surface meshes have been obtained. 
There, we randomly sampled from the complete set of available meshes.
This is the default case unless noted otherwise.

To gain more flexibility in the generation of particles with a specific form or even form distribution while maintaining characteristic features of sedimentary grains beyond the form, we applied an approach similar to \cite{zhao2020}.
As shown in \cref{fig:method:mesh_shape_scaling}, we preprocessed all available meshes, oriented along their principal axes, by scaling them such that their $L$, $I$, $S$ parameters became unity.
This step effectively removed all size and form related properties.
During the particle generation, we again randomly sampled from these meshes.
We then re-scaled the selected mesh in three dimensions with the vector $(\hat{L}, \hat{I}, \hat{S})$ to obtain the desired form and size.
This way, we preserved the possibly angular nature of the grain as much as possible.


\section{Simulation Setups and Procedures}
\label{sec:simulation_processes}

In this section, we introduce the setups, simulation processes and evaluation routines specific to our case.
We generally target random dense packings of fluvial sediment, modeled as if submerged and vertically deposited in water, and an accurate assessment of bulk porosity in two distinct simulation geometries.

\subsection{Domain Geometries} \label{sec:domain_geometries}

\begin{figure*}[t]
	\centering
	\begin{subfigure}{0.45\textwidth}
		\centering
		\begin{tikzpicture}
		\node[inner sep=0pt] (vis) at (0,0)
		{\includegraphics[trim=100 20 100 550, clip, height=5cm]{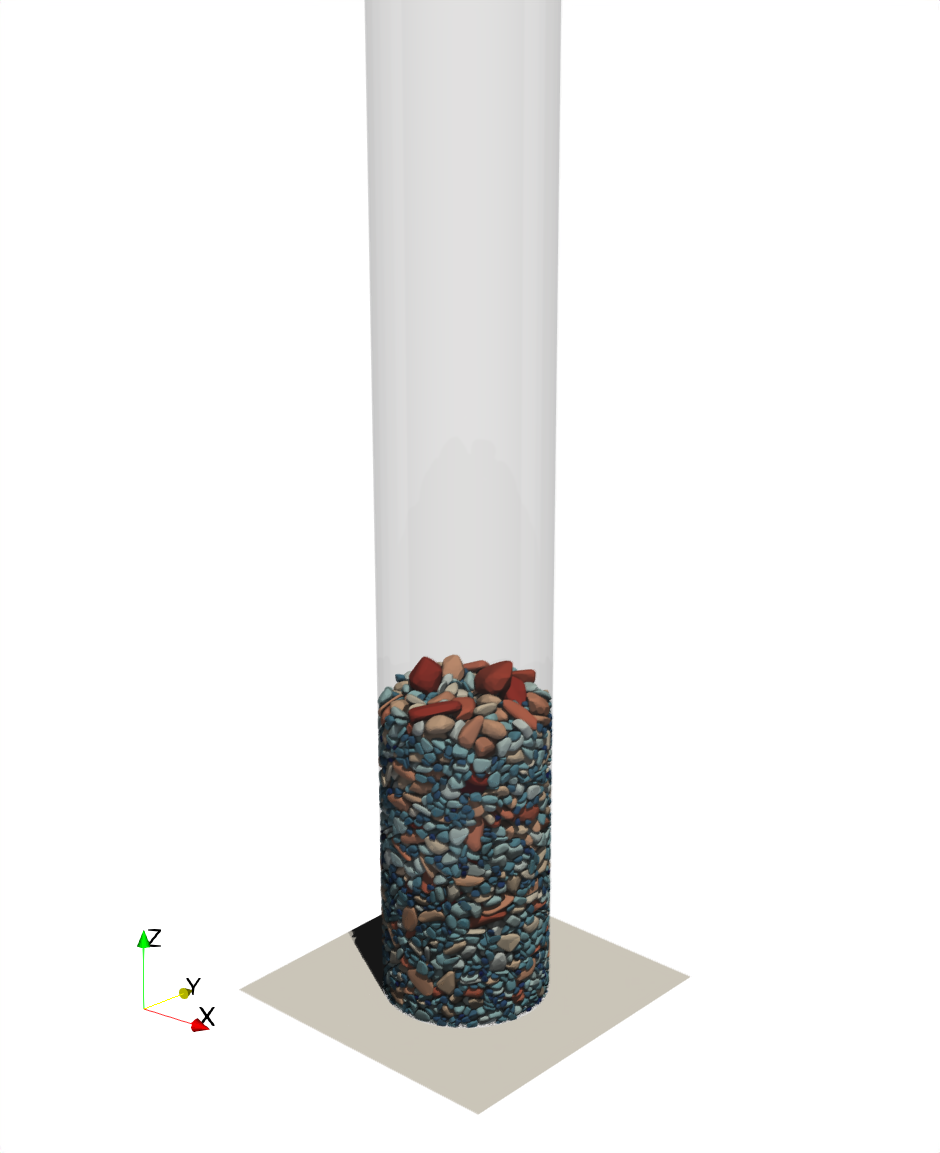}};
		\node[] (a) at ([xshift=2ex,yshift=-2ex]vis.north west) {\textbf{(a)}}; 
		\end{tikzpicture}
	\end{subfigure}~
	\begin{subfigure}{0.45\textwidth}
		\centering
		\begin{tikzpicture}
		\node[inner sep=0pt] (vis) at (0,0)
		{\includegraphics[trim=100 20 100 550, clip, height=5cm]{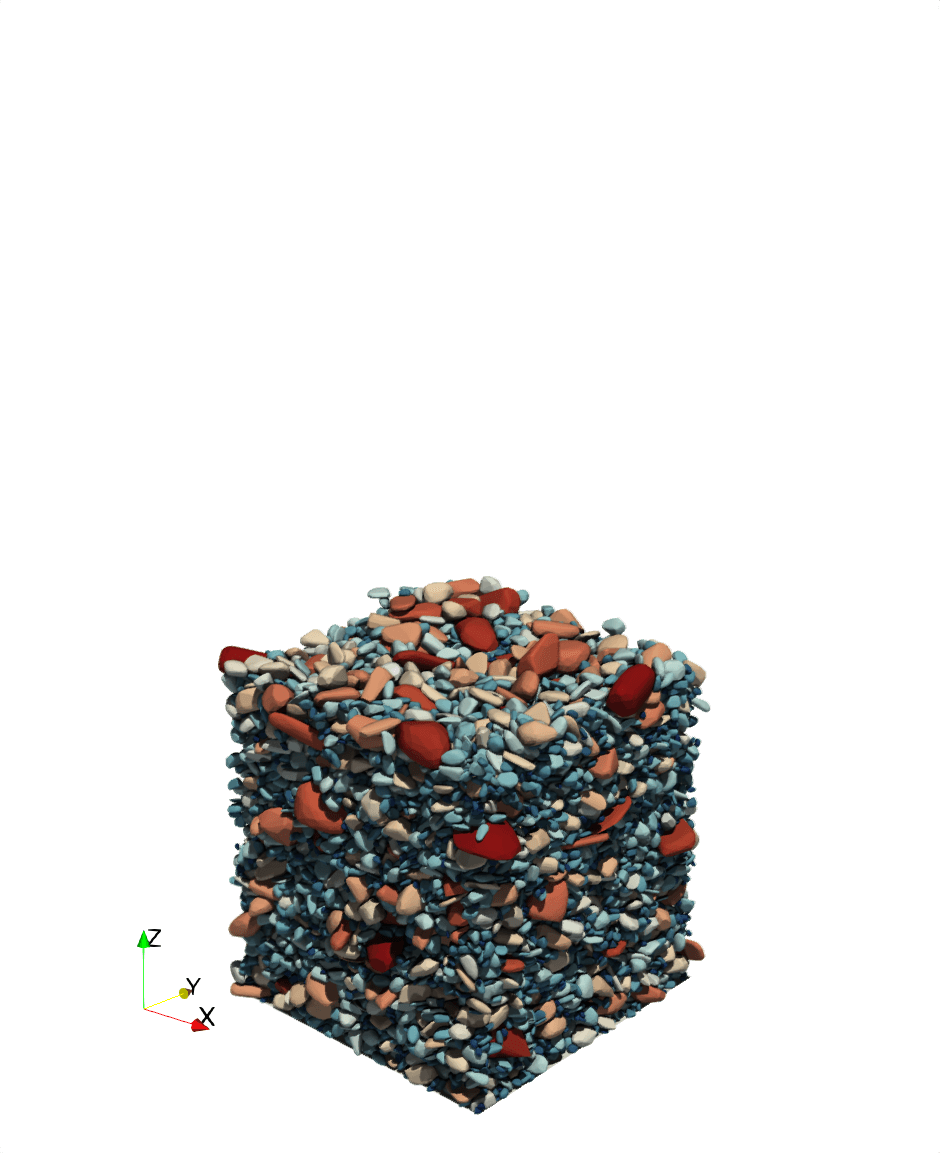}};
		\node[] (b) at ([xshift=2ex,yshift=-2ex]vis.north west) {\textbf{(b)}}; 
		\end{tikzpicture}
	\end{subfigure}
	\caption{Visualizations of the two domain geometries considered herein with exemplary particle packings, where a logarithmic color scale is used to depict the grain size. \textbf{a} cylinder confined, \textbf{b} horizontally periodic.} 
	\label{fig:setup_geometries}
\end{figure*}

In our studies, we considered two different domain setups, both featuring a bottom plane on top of which the particles formed the dense packing.
The first one, used in the calibration and validation studies in \cref{sec:validation}, was a cylindrical domain of width $L$ and height $H$ (see \cref{fig:setup_geometries}a). 
With this confined geometry, we aimed to reproduce the laboratory experiments from \cite{liang2015} as close as possible, where the same setup was used to determine the porosity value via the water displacement method.

It is well-known, however, that confining walls influence the packing structure in their vicinity and lead to increased porosity values \cite{zou1995,liu2000}.
Since our objective was to study bulk porosity, such a confined geometry would require large domain sizes to reduce this wall-effect.
Instead, we utilized another benefit of simulations and used a horizontally periodic domain (see \cref{fig:setup_geometries}b).
This second setup was applied for all our studies in \cref{sec:study_single,sec:study_mixture} where it modeled a small representative sample of the riverbed.
For this requirement to hold, the domain width $L$ had to be chosen large enough to avoid malicious influences of the regularity implied by the periodicity \cite{liu2000} which were established by sensitivity studies.

\subsection{Simulation Parameterization}

In all our simulations, we set the particle density to $\rho_p = 2650\,$kg/m$^3$, corresponding to quartz which is the dominant lithology in many rivers, like the Rhine.
We assumed that the grains were submerged in water with density $\rho_f = 1000\,$kg/m$^3$.
Together with a gravitational acceleration of $g = 9.81\,$m$^2$/s, an external force encompassing gravity and buoyancy of 
\begin{equation}
	\boldsymbol{F}_{p,i}^\mathit{ext} = - (\rho_p - \rho_f) V_{p,i} (0,0,g)^\top
\end{equation}
was acting on each grain in the vertical, here $z$, direction.

For the temporal integration of the equations of motion, \cref{eq:method:particle_translation,eq:method:particle_rotation}, we employed a time step size of $\Delta t = 5 \times 10^{-6}\,$s, similar to other packing studies of rock aggregates \cite{zhao2020}.
To parameterize our collision model, we made use of the coefficient of restitution $e_n$ and the collision time $T_c$ instead of specifying the stiffness and damping coefficients directly.
This approach has recently been used to successfully simulate dense polydisperse sphere packings \cite{rettinger2022} and offers direct control over the temporal resolution of single collision events.
Thus, the normal collision coefficients of \cref{eq:method:DEM_normalCollisionForce} are given as
\begin{equation}
k_n = \frac{m_{ij,\mathit{eff}}(\pi^2+\ln^2e_n)}{T_c^2},\quad d_n = -\frac{2{m_{ij,\mathit{eff}}\ln e_n}}{T_c},
\label{eq:DEM_normal_parameters}
\end{equation}
with the effective mass $m_{ij,\mathit{eff}} = \frac{m_{p,i}m_{p,j}}{m_{p,i}+m_{p,j}}$.
Here, we chose $e_n = 0.1$ and $T_c = 20\times 10^{-5}\,$s.
The relatively low value for the coefficient of restitution was used to mimic the damping behavior of the surrounding water, as e.g. also applied by \cite{ferdowsi2017}. 
The specific value for the collision time implied that each head-on collision is resolved within around 40 time steps, permitting enough time to have a smooth variation of the collision force and allow for stable simulations. 


In tangential direction, we followed \cite{thornton2013} and used a direct relation to the normal coefficients
\begin{equation}
k_t = \kappa_p k_n,\quad d_t = \sqrt{\kappa_p}d_n, 
\label{eq:DEM_tangential_parameters}
\end{equation}
which introduces the parameter
\begin{equation}
\kappa_p = \frac{2(1-\nu_p)}{2-\nu_p},
\end{equation}
with the Poisson's ratio $\nu_p$, a material property, for which we set $\nu_p=0.22$.
At last, the friction coefficient was here taken as $\mu_p = 0.5$, similar to other simulations \cite{schruff2018,gong2019,tong2015}.

\subsection{Simulation Process} \label{sec:simulation_process}

The generation process for dense random particle deposits featured three distinct phases, as briefly described next.

In the first phase, a certain number of particles was regularly generated until the desired total mass $m_\mathit{tot}$ of the packing was reached.
This sequential addition of grains can be seen as an efficient surrogate model for actual sediment deposition processes. 
There, we created particles inside the upper region of the computational domain, i.e., $z \in [0.6,1]H$, on a hexahedral grid with spacing $h_g$.
Given a large enough spacing, we could thus avoid huge initial overlaps of the particles which would otherwise destabilize the simulation immediately.
At each of the thus defined positions, we then selected the particle's size and shape according to the type of study and as explained in \cref{sec:method:generation}.
Additionally, we used a velocity $u_p^\mathit{init} = 1\,$m/s to initialize the translational and rotational velocity.
In particular, the vertical velocity was set to $u_{p,z} = - u_p^\mathit{init}$ while both horizontal components were sampled from a uniform distribution with the range $[-0.1, 0.1]u_p^\mathit{init}$. 
Similarly, the initial rotational velocities were obtained by sampling from the same interval and then dividing by the size $D$.
This measure effectively removed the regularity introduced by the hexahedral grid within a few time steps. 
The particles then settled due to the action of gravity and formed a dense packing on top of the bottom plane.
Once all particles of a thus generated sample had left the generation domain, a new set of particles was generated.

Throughout this generation phase and also beyond, we applied a horizontal shaking of the particles to encourage a dense rather than a loose packing \cite{an2008}.
We followed the approach of \cite{schruff2018} and imposed an additional external force in $x$ direction onto all particles, given as 
\begin{equation}
	\boldsymbol{F}_{p,i}^s(t) = m_{p,i} A_s \sin\left(\frac{2\pi}{T_s} t\right ) \frac{2\pi}{T_s}^2 (1,0,0)^\top,
\end{equation}
with shaking amplitude $A_s$ and period $T_s$. 
We noticed that it was important to apply this type of shaking directly from the beginning to avoid loose packings of the lower particle layers that cannot be compacted further once they are at rest.
Once all particles have been created, we continued the shaking for $t_s$ seconds.
As in \cite{schruff2018}, we used $A_s=3\times 10^{-4}\,$m and $T_s=0.025\,$s and determined $t_s$ via sensitivity studies. 
Due to the rather weak shaking strength, we have not observed a vertical size segregation of polydisperse setups, known as the chestnut effect. 
It thus complied with our goal to obtain a packing with an overall isotropic distribution of size and shape.

Once we switched off the shaking, the third, and final, phase of the simulation started where we actively damped the particle motion.
This procedure mimics the damping effect by the surrounding viscous fluid and dissipates the system's energy, accelerating the convergence to a steady, quiescent state of the packing.
Again following \cite{schruff2018}, we reduced the particles' translational and rotational velocities by multiplication with a damping factor $c_d$ in each time step, i.e.,
\begin{equation}
	\boldsymbol{u}_p(t) := c_d^{\Delta t} \boldsymbol{u}_p(t), \quad \boldsymbol{\omega}_p(t) := c_d^{\Delta t} \boldsymbol{\omega}_p(t).
\end{equation}
Here, we set $c_d = 10^{-3}$.


We assumed a converged simulation when the change in the vertical mass-average packing position, i.e., $\langle z \rangle = \sum_i m_{p,i} z_{p,i} / m_\mathit{tot}$, was below $10^{-5}\,$m for a time interval of $0.01\,$s.
This condition effectively checked if the compaction velocity was small enough.

\subsection{Porosity Evaluation}
\label{sec:porosity_evaluation}

\begin{figure*}[t]
	\centering
	\begin{tikzpicture}
	\node[inner sep=0pt] (vis) at (0,0)
	{\includegraphics[trim=200 190 100 300, clip, height=5cm]{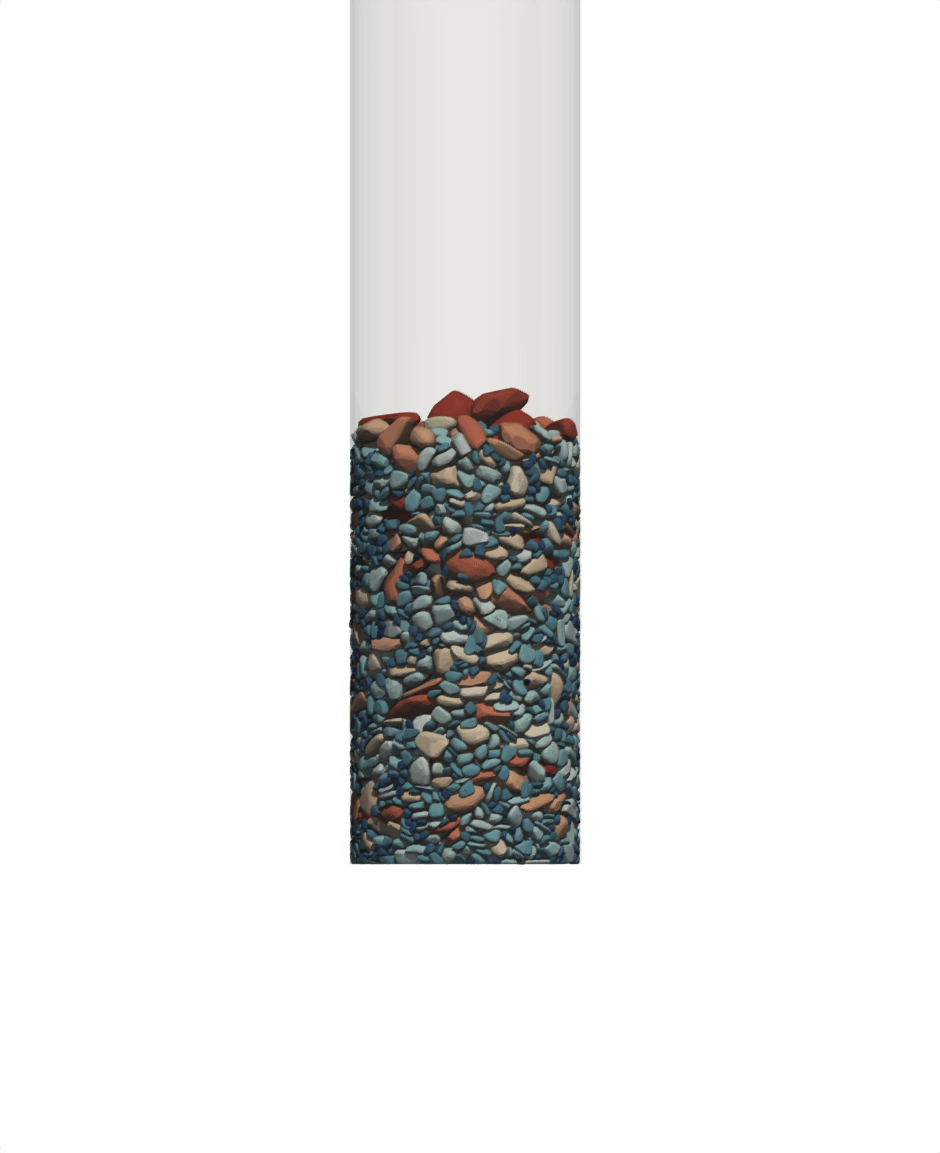}};
	\node[inner sep=0pt] (vis) at (8.7,0)
	{\includegraphics[trim=100 180 100 240, clip, height=5cm]{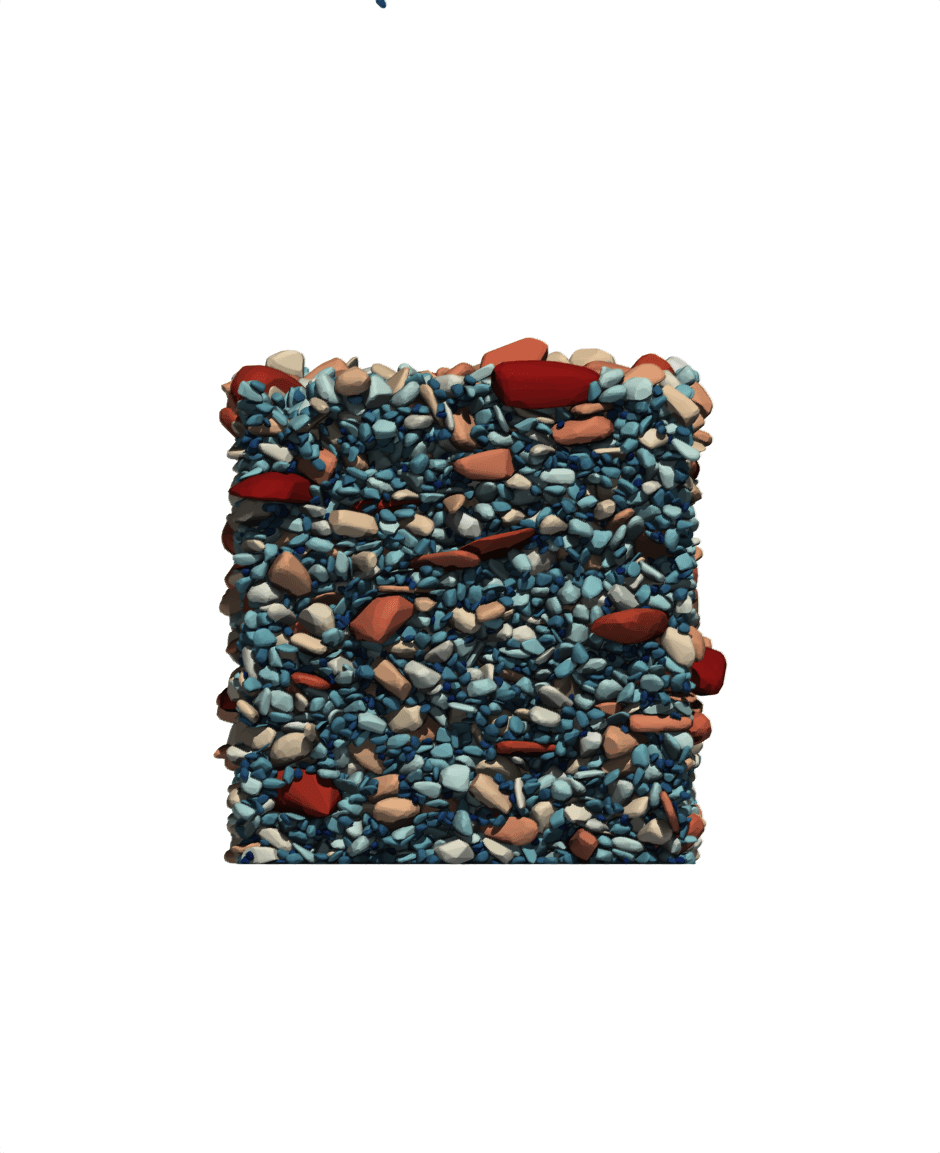}};
	\node[inner sep=0pt] (vis) at (3.7,0)
	{\includegraphics[height=5cm]{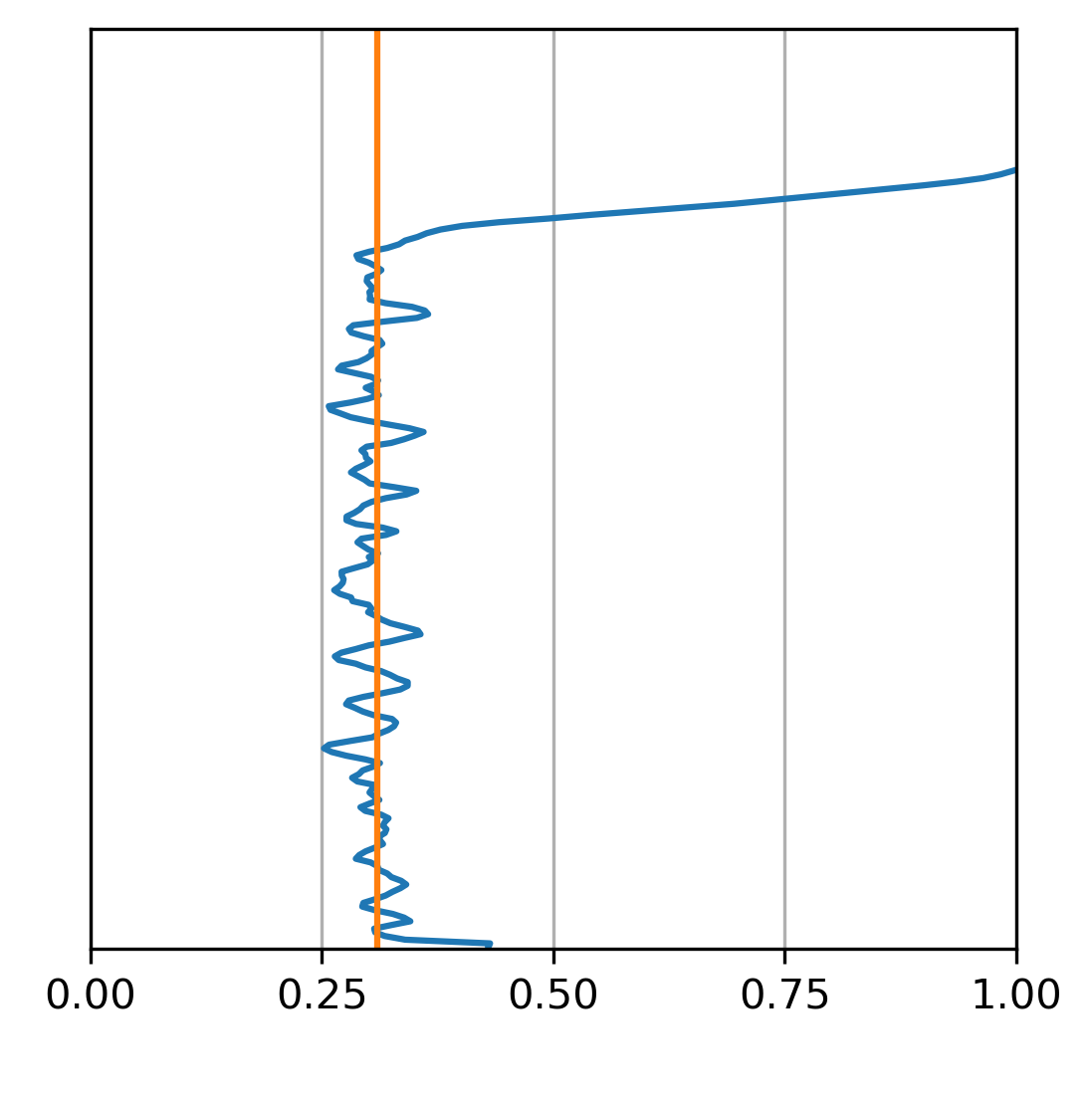}};
	
	\fill[fill=color3, fill opacity=0.1] (-1.4,-1.78) rectangle ++(7.25,3.3);
	\fill[fill=color5, fill opacity=0.1] (1.65,-1.28) rectangle ++(8.9,2.28);
	
	\draw[thick, color3] (-1.4,-1.78) -- ++(7.25,0) node[pos=1,right,color=black]{$z_-^\mathit{cd}$};
	\draw[thick, color3] (-1.4,1.52) -- ++(7.25,0) node[pos=1,right,color=black]{$z_+^\mathit{cd}$};
	\draw[thick, color5] (1.65,-1.28) -- ++(8.9,0) node[pos=0,left,color=black]{$z_-^\mathit{pd}$};
	\draw[thick, color5] (1.65,1) -- ++(8.9,0) node[pos=0,left,color=black]{$z_+^\mathit{pd}$};
	\draw[thick, black, latex-latex] (5,1) -- ++(0,0.5) node[pos=0.5,right]{$h^\mathit{pd}$};
	\draw[thick, black, latex-latex] (5,-1.78) -- ++(0,0.5) node[pos=0.5,right]{$h^\mathit{pd}$};
	\node[] at (3.7,-2.3) {porosity $n$};
	\node[rotate=90] at (1.4,-0.1) {height $z$};
	\node[] at (2.7,2) {$\langle n\rangle$};
	\end{tikzpicture}
	\caption{Sketch of the porosity evaluation based on the vertical porosity profile, including the definition of the evaluation intervals $[z_-^\mathit{cd},z_+^\mathit{cd}]$ for the cylindrical domain (left part) and $[z_-^\mathit{pd},z_+^\mathit{pd}]$ for the horizontally periodic domain (right part) which are depicted as shaded areas.}
	\label{fig:porosity_evaluation}
\end{figure*}
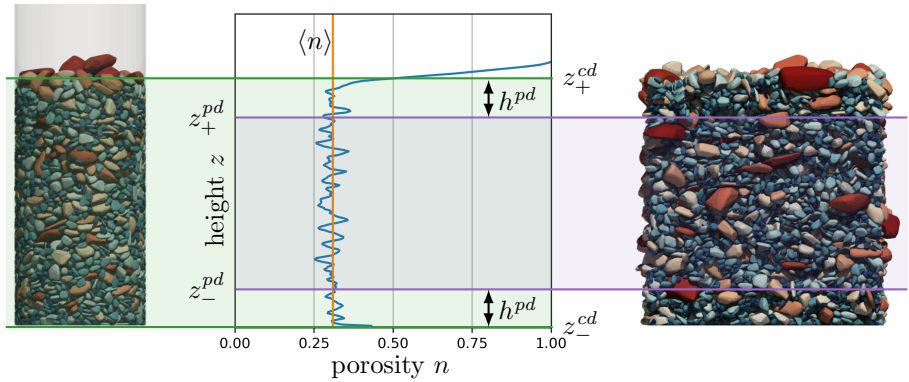

As the primary result, we are interested in the bulk porosity of the packing which was evaluated in a post-processing step after convergence of the simulation.
Since the full three-dimensional information about the pore sizes and location was thus not required, we first extracted a vertical porosity profile.
Such a profile can be obtained conveniently by replacing each particle by a volume-equivalent sphere at the same position and following a procedure as in \cite{rettinger2022}.
We, thus, subdivided the computational domain into horizontal slices of height $h_p$, stacked vertically upon each other.
Here, we used $h_p=10^{-3}\,$m.
The intersection of a sphere with such a slice is given by a spherical segment, whose intersection volume can be computed analytically.
Accumulating these volumes for each particle and each slice, we obtained the total particle volume per slice. 
The major advantage of this approach is that the particle volume is exact and without a sampling error, as would be introduced by voxelization approaches, see e.g. \cite{schruff2018}.
Subtracting this value from the also known slice volume and dividing again by the latter finally yielded $n(z)$, i.e., porosity as a function of height.
From this profile, the bulk porosity can be obtained by averaging over the bulk region via 
\begin{equation}
	\langle  n \rangle  = \frac{1}{z_+ - z_-}\int_{z_-}^{z_+} n(z) \text{ d} z
\end{equation} 
where $z_-$ and $z_+$ are the lower and upper limits of the evaluation interval.
The definition of these limits resembles the treatment of the wall-effect, imposed by the bottom plane, and the free surface at the top.
It differed for the two setups considered here and led to the evaluation intervals $[z_-^\mathit{cd},z_+^\mathit{cd}]$ for the cylindrical domain, and $[z_-^\mathit{pd},z_+^\mathit{pd}]$ for the horizontally periodic one.
This procedure is sketched in \cref{fig:porosity_evaluation}.

For the cylindrical domain, we set $z_-^\mathit{cd}=0\,$m since the corresponding laboratory experiments used the water-displacement method and thus, the extracted porosity values also included the wall-effect at the bottom.
On the upper part, the porosity profile rapidly increases from the bulk value to its maximum value of 1.
There, we aimed to exclude the effect of outliers, i.e., single particles lying on top of the otherwise dense packing. 
In the actual laboratory experiments, the upper surface of the packing was manually made even to achieve this.
In the simulations, we chose $z_+^\mathit{cd}$ such that $n(z_+^\mathit{cd})=n^\mathit{cut-off} = 0.5$.

For the horizontally periodic setup, we intended to eliminate effects from the lower and upper boundary all together to extract the actual bulk porosity.
Here, we followed the approach for the cylindrical domain and additionally cut away a region of height $h^\mathit{pd}$ from both ends, such that  $z_-^\mathit{pd} = h^\mathit{pd}$, and $z_+^\mathit{pd} = z_+^\mathit{cd} - h^\mathit{pd}$.

To avoid cluttered notation, we will report the resulting bed porosity as $n$ instead of $\langle n \rangle$.


\section{Calibration and Validation in Confined Domain}
\label{sec:validation}

In this section, we present packing simulations that virtually reproduced the experiments from \cite{liang2015} in a cylindrical domain.
Via a sensitivity study, we determined the influence of specific process parameters and calibrated those to the experiments.
Taking the experimental porosity measurements as a reference, we finally validated our simulation approach and demonstrated its usefulness for predictive packing simulations.

\subsection{Simulation Description}

\begin{table*}[t]
	\centering
	\begin{tabular}{l|ccccccc|cc}
		\toprule
		 & 2.8 - & 4 -& 5.6 - & 8 - & 11.2 - & 16 - & 22.4 - &  & \\
		Case & 4 & 5.6& 8 & 11.2 & 16 & 22.4 & 31.5 & $n^\mathit{exp}$ & $h_g$ (mm) \\ \midrule
		U1 & .0 & .0 & 1. & .0 & .0 & .0 & .0 & 0.371 $\pm$ 0.001 & 16 \\
		U3 & .0 & .0 & .21 & .58 & .21 & .0 & .0 & 0.356 $\pm$ 0.003 & 22 \\
		U5 & .0 & .06 & .24 & .4 & .24 & .06 & .0 & 0.344 $\pm$ 0.002 & 26 \\
		U7 & .04 & .11 & .22 & .26 & .22 & .11 & .04 & 0.315 $\pm$ 0.002 & 22 \\
		B30 & .08 & .13 & .08 & .06 & .18 & .29 & .18 & 0.270 $\pm$ 0.003 & 20 \\
		B50 & .13 & .21 & .13 & .06 & .13 & .21 & .13 & 0.289 $\pm$ 0.005 & 20 \\
		B70 & .18 & .29 & .18 & .06 & .08 & .13 & .08 & 0.299 $\pm$ 0.002 & 18 \\
		\bottomrule
	\end{tabular}
\caption{Mass-based size fractions and related parameters for the simulation studies. The sieve sizes are taken from the laboratory experiments in \cite{liang2015} and given in mm. The case names resemble whether the packing is unimodal (U), with the number of size fractions given as the numeral, or bimodal, where the numeral indicates the mass percentage of fine grains. $n^\mathit{exp}$ is the average porosity together with its standard deviation obtained from the laboratory experiments. The last column denotes the generation spacing in the simulations.}
\label{tab:setup_experiments}
\end{table*}

We considered the setup with a cylindrical domain, as described in \cref{sec:domain_geometries} and also applied in the laboratory experiments from~\cite{liang2015}.
In the latter, gravel-sized fluvial grains from the Rhine have been dropped into this domain, that was partially filled with water and of size $L=0.104\,$m and $H = 1\,$m.
The total grain mass was $m_\mathit{tot} = 3$ kg and seven different size distributions have been used.
Each experiment has been carried out twice with the same sample to account for the possibly stochastic nature of the deposition process.
These size distributions and the measured porosity values are summarized in \cref{tab:setup_experiments}.
During the process, some tapping in the side walls was applied to remove trapped air bubbles that also further compacted the packing. 
No further quantitative information about this process is available.
Here, we assumed that our slight shaking as described in \cref{sec:simulation_process} is qualitatively comparable.
The initial spacings $h_g$ applied for newly generated particles depended on the grain size distribution in order to avoid excessive initial surface overlaps between neighboring particles but permit a large number of particles per generation step.
These spacings are also reported in \cref{tab:setup_experiments}.

For this comparably small sample mass, we noted that especially the largest size fraction was typically only represented by a few individual grains.
These large particles, however, have a major influence on the actually generated size distribution and the obtained porosity.
Consequently, the randomness of the sampling during particle creation, for the size as well as for the actual grain shape as described in \cref{sec:method:generation}, was expected to have a certain effect on our simulation results.
In order to increase the robustness of our results, we thus simulated each case three times and ensured that the random number generators were initialized with different seeds each time, producing distinct packings. 

\subsection{Calibration Study}
\label{sec:val:calibration}

To study the influence of mesh resolution and shaking time, we carried out sensitivity studies to investigate the respective parameter space and identify suitable settings.
We picked the cases U7 and B50 as they offer the broadest span of particle sizes and contain a unimodal and a bimodal case.
Consequently, we expected the result to carry over to the other size distributions as well and could keep the computational effort low.

\subsubsection{Effect of Grain Surface Mesh Resolution}

\begin{table*}[t]
	\centering
	\begin{tabular}{ll|rrrrr}
		\toprule
		Case & & no simplification & $n_F\leq300$ & $n_F=200$ &  $n_F=100$ & $n_F = 50$   \\
		& volume change & - &  $-0.2\%$ & $-0.4\%$ & $-1.3\%$ & $-3.6\%$ \\\midrule
		U7 & $n$ & 0.309 & 0.310 & 0.315 & 0.319 & 0.329 \\
		 & runtime (h) & 2.97 & 1.92 & 1.98 & 1.69 & 1.48 \\\addlinespace
		B50 & $n$ & 0.274 & 0.278 & 0.284 & 0.289 & 0.294\\
		& runtime (h) & 8.25 & 5.72 & 4.97 & 4.29 & 4.04\\
		\bottomrule
	\end{tabular}
	\caption{Effect of mesh resolution on porosity and simulation runtime of packings in a cylindrical domain. The second row reports the average relative volume changes compared to the unsimplified mesh, averaged over all 63 meshes.}
	\label{tab:val:sensitivity_mesh_res}
\end{table*}

As presented in \cref{sec:form_meshes}, the meshes originated from differently sized sediment grains that have been scanned with a certain spatial resolution.
The surface meshes for small particles thus featured significantly less triangles than for larger grains, with the number of triangles ranging from 220 to 2620.
In order to directly compare simulations with different meshes, a similar number of triangles would be preferable.
Additionally, the collision detection for meshes with higher resolution 
takes more time, increasing the computational cost.
Therefore, there was a strong incentive to reduce the number of triangular faces $n_F$ that describe the particle shape, ideally such that the most relevant features were still captured and all meshes had roughly the same resolution.
We here made use of the simplification functionality provided by the PyMeshLab library \cite{pymeshlab} and study the influence of such a pre-processing step on porosity.

Specifically, we applied the routine \texttt{simplification quadric edge collapse decimation}, and required a significantly reduced number of faces between 50 and 300.
For the latter, original meshes with already less than 300 faces remained unchanged by this procedure.
From the average relative volume changes reported in \cref{tab:val:sensitivity_mesh_res}, we see that these mesh simplifications generally reduced the grain volume by at most $4$\%.
For the packing simulations, we set $t_s = 2$ s and report the obtained porosity and simulation runtime, average over the three distinct realizations, in the same table.
For both cases, U7 and B50, we noticed an increase in porosity if the mesh resolution was decreased, with an absolute difference of up to 0.02 when comparing the original and the strongly simplified case.
We expect this to be a result of the more angular particles in the latter that gave rise to larger pore spaces than the smoother meshes.
At the same time, the simulation runtimes dropped by roughly a factor of 2 in the two limiting cases.
Since almost the same porosity values were obtained for the case of $n_F \leq 300$ with a runtime reduction of around 1/3, we used this degree of mesh simplification for all upcoming simulations.

\subsubsection{Effect of Shaking Duration}

\begin{table*}[t]
	\centering
	\begin{tabular}{l|cccc}
		\toprule
		Case & $0\,$s & $1\,$s & $2\,$s & $3\,$s \\\midrule
		U7 & $0.323 \pm 0.001$ & $0.314 \pm 0.001$ & $0.309 \pm 0.001$ & $0.309 \pm 0.002$ \\
		B50 & $0.284 \pm 0.004$ & $0.280 \pm 0.003$ & $0.280 \pm 0.004$ & $0.277 \pm 0.004$\\  
		\bottomrule
	\end{tabular}
	\caption{Effect of shaking duration $t_s$ on porosity of packings in a cylindrical domain.
	}
\label{tab:val:sensitivity_shaking}
\end{table*}

Next, we investigated the effect of the shaking procedure on porosity, parameterized by the shaking duration $t_s$.
It was introduced in \cref{sec:simulation_process} as a measure for an adequate compaction to ensure a dense packing but is not directly present in the experiments.
Generally, shorter shaking times are preferable as they result in shorter simulation runtimes and avoid a vertical size segregation of the grains.

For this study, we varied the shaking duration between 0 and $3\,$s of continued shaking after the generation of all particles was finished.
The result of this study is reported in \cref{tab:val:sensitivity_shaking}.
The effect was strongest for the case U7, where a significant reduction of porosity could be observed when comparing the cases of no continued shaking to the others.
Shaking longer than 1$\,$s further decreased porosity but the effect is weaker.
For B50, we could also observe a reduction in porosity, although to an overall smaller extent.
We note that in this case, we aimed for an adequate representation of the laboratory tapping on the outside of the cylinder rather than generating packings that are as dense as possible.
Comparing with the experimentally measured porosity values, see \cref{tab:setup_experiments}, we decided to use $t_s = 1\,$s for the following validation studies.





\subsection{Validation Study}
\label{sec:val:validation}

\begin{table*}[t]
	\centering
	\begin{tabular}{l|ccccrr}
		\toprule 
		Case & run 1 & run 2 & run 3 & mean $\pm$ std. dev. & MAE & $\mathcal{E}$ (\%) \\\midrule
U1  &  0.382 & 0.381 & 0.386 & 0.383 $\pm$ 0.002 & 0.012 & 3.29 \\
U3  &  0.367 & 0.369 & 0.372 & 0.369 $\pm$ 0.002 & 0.013 & 3.76 \\
U5  &  0.336 & 0.333 & 0.332 & 0.333 $\pm$ 0.002 & -0.011 &-3.07 \\
U7  &  0.315 & 0.313 & 0.314 & 0.314 $\pm$ 0.001 & -0.001 & -0.41 \\
B30  &  0.284 & 0.290 & 0.283 & 0.286 $\pm$ 0.003 & 0.016 & 5.86 \\
B50  &  0.283 & 0.276 & 0.281 & 0.280 $\pm$ 0.003 & -0.009 & -3.12 \\
B70  &  0.290 & 0.291 & 0.290 & 0.290 $\pm$ 0.001 & -0.009 & -2.99 \\
\bottomrule
	\end{tabular}
	\caption{Simulated porosity values of the setups given in \cref{tab:setup_experiments}, together with their mean absolute error (MAE) and relative error $\mathcal{E}$, both evaluated with respect to the experimental porosity values.}
	\label{tab:val:results}
\end{table*}

To finally validate our simulation approach for the virtual study of gravel-sized grain packings, we used all seven grain size distributions given in \cref{tab:setup_experiments}.
The results of the three individual realizations, their average and standard deviation, and the corresponding relative error calculated based on $n^\mathit{exp}$ are given in \cref{tab:val:results}.
Like in the experiments, we observed a reduction in porosity in the unimodal cases when more size fractions were included. 
This behavior is qualitatively in line with the experiments and other findings for spherical particles \cite{brouwers2014}.
The absolute and relative errors were at most $0.013$ and $3.76\,\%$, respectively, attesting a very good agreement between simulations and experiments.
Similarly, the cases B50 and B70 agreed well with the experiments.
Only the case B30 exhibited a notably higher error as porosity there increased in comparison to B50, different from the experiments where a further reduction has been observed.
Further studies where other simulation parameters were altered selectively confirmed this trend of the simulations.
Furthermore, the additional experimental data with glass spheres or grains from another river reported by \cite{liang2015} had also exhibited the porosity increase for case B30 instead of an decrease. 
Therefore, there could be a certain operational or process-related bias present in the experimental data for that case that might explain this deviation. 

Generally, we note that the obtained results should be seen in regard of other possible sources of deviations.
As already mentioned, the shaking procedure was different in simulations and experiments, and the findings from the previous calibration study showed that such packings are sensitive to the degree of shaking.
Moreover, the experiments naturally featured grains with distinct shapes whereas the simulation sampled only from a limited number of shapes.
The agreement that we achieved despite these factors with our simulation method can be further appreciated as it was able to yield considerably improved results in comparison to the stochastic packing algorithm applied in \cite{liang2015}.
There, relative errors between 17 and 42$\,\%$ have been reported, compared to the errors between 0 and 6$\,\%$ in the present study. 
We thus conclude from these results that our simulation approach is suitable to accurately predict the porosity of sedimentary grain packings.

\section{Porosity of Single Shape Packings}
\label{sec:study_single}

To allow a systematic study of the influence of grain form on porosity, we first considered systems that featured a single shape, and by implication a single form, for all grains in the following. 
The goal was to develop an understanding of the most important form factors with respect to porosity based on extensive data obtained from simulations.
Finally, a predictive model was constructed for single shape packings.

\subsection{Simulation Description}

Here, all generated packings used for the model development were made up of a single shape, as given by one of the 63 available surface meshes.
Such a study of packings with single, but complex shapes benefits greatly from the possibility of a simulation-based approach and would not, or only with a disproportionally high effort, be feasible with laboratory experiments.
Due to the explicit focus on shape-effect, we primarily made use of the unimodal size distribution with a single size fraction, i.e., case U1 from \cref{tab:setup_experiments}.
However, since it is known that the size distribution has a strong effect on porosity, we additionally considered the cases U7 and B50 from the same table to explore the generality and robustness of the obtained findings.

To eliminate the effect of the bounding walls, present in the studies with a cylindrical domain in \cref{sec:validation}, we used the horizontally periodic domain described in \cref{sec:domain_geometries}.
To additionally get rid of the effect of bottom and top wall in order to just assess the porosity of the bulk region, we cut away the size of the maximum sieve fraction, i.e., $h^\mathit{pd} = 31.5\,$mm, in all cases as defined in \cref{sec:porosity_evaluation}.

\subsection{Sensitivity study}

We first carried out a sensitivity study to find a suitable horizontal domain size that avoids effects of periodicity-introduced regularity and guarantees large enough particle numbers per size fraction.
Otherwise, simulations where those requirements are not met could not be considered a valid representation of such a packing.
Furthermore, we again investigated the effect of the shaking duration.
This time, however, with the goal to obtain dense packings.

As in \cref{sec:val:calibration}, we used the size distributions U7 and B50 from \cref{tab:setup_experiments} for these studies.
Since U7 features the smallest fraction of particles from the largest size fraction, we expected that the largest sample sizes are required for this case.
In all cases, we again used three different realizations to generate the packings.
Since those calibration studies also served as a basis for the form mixture studies in \cref{sec:study_mixture}, we sampled the particles' shape randomly from all available meshes when generating the particles, as done in \cref{sec:validation}.
Thus, in contrast to the other packings considered in the present section, we purposely used shape mixtures here.


\subsubsection{Effect of Horizontal Domain Size}

\begin{table}[t]
	\centering
	\begin{tabular}{l|ccc}
		\toprule
		 & $0.1\,$m & $0.2\,$m & $0.3\,$m  \\
		Case & $4\,$kg & $16\,$kg & $36\,$kg \\\midrule
		U7 & 0.291 $\pm$ 0.001 & 0.291 $\pm$ 0.002 & 0.291 $\pm$ 0.001 \\
		B50 & 0.246 $\pm$ 0.005 & 0.245 $\pm$ 0.003  & 0.251 $\pm$ 0.000 \\ 
		\bottomrule 		
	\end{tabular}
	\caption{Effect of horizontal domain size $L$, with the respectively increased sample mass $m_\mathit{tot}$, on porosity in the horizontally periodic setup.}
	\label{tab:ss:sensitivity_domain_size}
\end{table}

For these studies, we set the shaking duration to $t_s = 1\,$s and varied the horizontal domain size $L \in \{0.1, 0.2, 0.3\}\,$m.
To maintain roughly same packing heights in all cases, we increased the sample mass accordingly.
This way, the number of particles for case U7 varied from around 6000 to 55\,000 while B50 featured 11\,000 to 100\,000 particles.
The obtained porosity values are stated in \cref{tab:ss:sensitivity_domain_size}.
A first immediate outcome was that these values were significantly below the ones found for the confined geometry in the previous section since the porosity-increasing effect of the horizontally bounding walls was avoid.
Regarding the effect of the domain size, no influence on porosity could be observed for the case U7.
Also for B50, the effect of increasing the domain size was hardly visible but the standard deviation between the three runs got reduced.
As a compromise between ensuring that enough particles of the largest size fraction are present and the computational effort, we decided to use $L=0.2\,$m with $m_\mathit{tot} = 16\,$kg.
For the here considered maximum grain size of $D_\mathit{max} = 31.5\,$mm, $L/D_\mathit{max} \approx 6.35$ and the sample mass was well-above different available recommendations, as summarized by \cite{seitz2018}. 

\subsubsection{Effect of Shaking Duration}

\begin{table}[t]
	\centering
	\begin{tabular}{l|cccccc}
		\toprule
		Case & $1\,$s & $2\,$s & $3\,$s & $4\,$s & $6\,$s & $8\,$s \\\midrule
		U7 & 0.291 & 0.287  & 0.285 & 0.283 & 0.282 & 0.281  \\
		B50 & 0.245 & 0.241 & 0.241 & 0.243 & 0.242 & 0.241  \\
		\bottomrule
	\end{tabular}
	\caption{Effect of shaking duration $t_s$ on porosity in the horizontally periodic setup.}
	\label{tab:ss:sensitivity_shaking_duration}
\end{table}

The influence of the additional shaking duration on porosity is shown in \cref{tab:ss:sensitivity_shaking_duration}.
Similarly to the confined setup, longer shaking times generally resulted in denser packings.
For the wide unimodal size distribution of case U7, this effect was more pronounced and saturated at around $t_s=6\,$s.
The bimodal case B50 exhibited less influence of the shaking.
Note, however, that $t_s$ only quantified the duration of shaking applied after the continuous generation of particles was completed and shaking was nevertheless applied during the whole generation process.
We also noticed that the rearrangement processes during shaking were seemingly beneficial for the overall convergence behavior of the packing, i.e., the packing came to rest more easily.
For all upcoming studies, we used $t_s=6\,$s, again as a compromise regarding computational effort.

\subsection{Porosity Model for Single Form Packings}
\label{sec:ss:porosity_model}


Next, we considered packings of particles with only a single shape, prescribed by one of the 63 geometries.
Since those available geometries cover a wide range of form factors, see \cref{sec:form_meshes}, we aimed to develop a model that adequately represents the relation between two specifically selected form factors and porosity based on the simulation results.

\subsubsection{Description}

Generally, data-driven model development offers a huge variety of approaches \cite{kuhn2013}.
As discussed in \cref{sec:shape:correlation}, the different considered form factors, that are also referred to as features in the context of data-driven modeling, are highly correlated.
However, many of those data regression approaches require, or at least favor, independent features in order to be applied successfully.
Additionally, we wanted to maintain interpretability of the developed model rather than constructing a black-box-like model.
This property is also favorable for a general applicability of the model as a clear functional correlation is easier to handle than a complexly structured model with various parameters, like a neural net.
These requirements thus go hand in hand with keeping the model as simple as possible, restricting the analysis to linear dependencies between porosity and the features.

The 13 form factors we considered here, see \cref{tab:form_factors}, are dimensionless ratios and, thus, the original choice of denominator and numerator was in principle arbitrary.
For that reason, we also added the inverse of each form factor $X$, denoted by $X^{-1}$, to the set of features.
In total, we thus considered 26 features.
In addition to the uniform case U1, we also used the size distributions U7 and B50 to test the generalizability of the result to a broader unimodal distribution and a bimodal one.
Still, for brevity, the focus was primarily on the effect of form on porosity and we divert a quantification of the effect of the size distributions to future studies.

The packing simulations used the values obtained by the sensitivity study, i.e., $t_s=6\,$s and $L=0.2\,$m with $m_\mathit{tot} = 16\,$kg.
They were executed using 400 processes.
Since the variability between different random realizations of the same setup were generally found to be rather low, we only used a single one per chosen geometry which also kept the computational cost in bounds.

\subsubsection{Extraction of Primary Form Factor}

\begin{figure*}[t]
	\centering
	\input{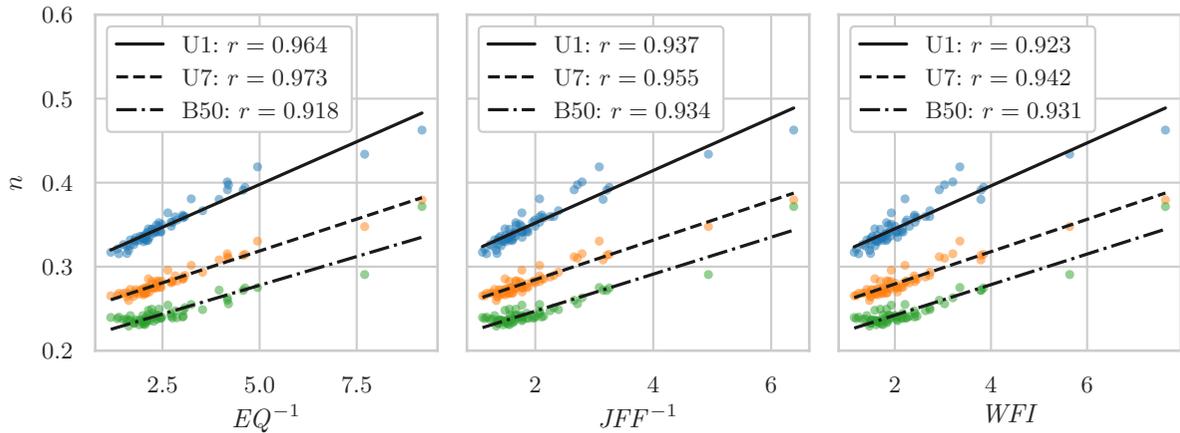}
	\caption{Three form factors that are correlated the most with porosity for packings with a single shape. The straight lines are the implied linear models, for which the correlation coefficient is given in the legend.
	}
	\label{fig:ss:formfactor_porosity_correlation}
\end{figure*}

The actual analysis of the data and the model development was done in a two step procedure to obey the requirements formulated above.
At first, we studied the correlation of each form factor with porosity for the set of 63 packings, individually for each size distribution.
This way, we could extract the single form factor that exhibited the most pronounced effect on porosity.
The correlation metric was given by Pearson's correlation coefficient to check for a linear relationship, which implicitly derived a linear model between this form factor and porosity.
This analysis is shown in \cref{fig:ss:formfactor_porosity_correlation} for the three form factors with the, in absolute values, largest correlation values.
They were given as the inverse equancy, the inverse of the Janke form factor, and the Wentworth flatness index.
The actual porosity values, and consequently the linear models, ware clearly affected by the size distribution.
This correlation ordering was found for both cases U1 and U7, whereas $\mathit{JFF}^{-1}$ was the highest-correlating for case B50.
From the definition of these form factors we noted that all three have in common that they feature ratios of the longest axis $L$ over the shortest axis $S$.
Based on these findings, we chose the inverse equancy, i.e., $L/S$, as the first model feature as it overall scored the largest correlation values and also is the simplest of the three candidates.


\subsubsection{Porosity Model Based on Two Form Factors}

\begin{table*}[t]
	\centering
	\setlength{\tabcolsep}{4pt} 
	\begin{tabular}{l|*{14}{c}}
		\toprule
		Case & \rotatebox[origin=c]{90}{$\mathit{FL}$} & \rotatebox[origin=c]{90}{$\mathit{EL}$} & \rotatebox[origin=c]{90}{$\mathit{EL}^{-1}$} & \rotatebox[origin=c]{90}{$\mathit{EQ}^{-1}$} & \rotatebox[origin=c]{90}{$\mathit{ASF}$} & \rotatebox[origin=c]{90}{$\mathit{ASF}^{-1}$} & \rotatebox[origin=c]{90}{$\mathit{IFI}$} & \rotatebox[origin=c]{90}{$\mathit{IFI}^{-1}$} & \rotatebox[origin=c]{90}{$\mathit{IRI}$} & \rotatebox[origin=c]{90}{$\mathit{IRI}^{-1}$} & \rotatebox[origin=c]{90}{$\mathit{OPI}$} & \rotatebox[origin=c]{90}{$\mathit{OPI}^{-1}$} & \rotatebox[origin=c]{90}{$\mathit{DRI}$} & \rotatebox[origin=c]{90}{$\mathit{DRI}^{-1}$} \\\midrule
		U1 & 5.96 & 5.26 & \textbf{4.66} & 5.74 & 5.40 & 5.58 & 5.18 & 5.56 & \textbf{4.80} & \textbf{4.19} & 5.48 & 6.96 & 5.71 & 5.87 \\
		U7 & 3.90 & 3.98 & \textbf{3.73} & 4.09 & \textbf{3.55} & 3.93 & \textbf{3.65} & 3.88 & 4.14 & 3.96 & 3.93 & 4.97 & 3.92 & 4.11 \\
		B50 & \textbf{6.02} & 6.16 & 6.46 & 6.12 & 6.21 & 6.17 & 6.40 & 6.29 & \textbf{5.69} & 6.28 & 6.32 & 6.30 & 6.25 & \textbf{5.86} \\
		\bottomrule
	\end{tabular}
	\caption{Mean absolute error ($\times 10^{-3}$) of predicted porosity to the one obtained from the simulation for the different combinations of $\mathit{EQ}^{-1}$ with any other form factor. The three smallest ones for each case are given in bold.}
	\label{tab:ss:two_factors_result}
\end{table*}

Next step, we intended to find the second form factor that augments the predictive quality of the first one and thus completes the form description and the porosity model, see \cref{sec:shape:correlation}.
We refrained from simply choosing the second-best feature from the previous analysis as it might be highly correlated to the inverse equancy and would thus not add additional information. 
Therefore, we first removed all features from the feature set that were highly correlated to $EQ^{-1}$ which we here defined by an absolute pair-wise correlation coefficient above 0.8.
In our case, this procedure effectively removed 12 features from the set.
With the remaining ones, we carried out a bivariate linear regression against the observed porosity values, always considering pairs of the inverse equancy and any other form factor.
We evaluated the performance of this approach in terms of the medium absolute error of a 10-fold cross validation as a robust measure.
These errors are reported in \cref{tab:ss:two_factors_result}, again for the three different size distributions.
The main behavior was already well-predicted by the first form factors alone, as visible in the $\mathit{EQ}^{-1}$ column.
Consequently, we observed overall only minor differences between the investigated feature pairs.
For the cases U1 and U7, however, the combination with the inverse elongation ($\mathit{EL}^{-1}$) ranked among the three smallest values.
Therefore, we decided to take this form factor as the second model parameter.
This combination was less ideal for the bimodal case B50, but there the error was still as low as around $6\times 10^{-3}$, almost independent of the chosen feature pair.

\begin{figure*}[t]
	\centering
	\input{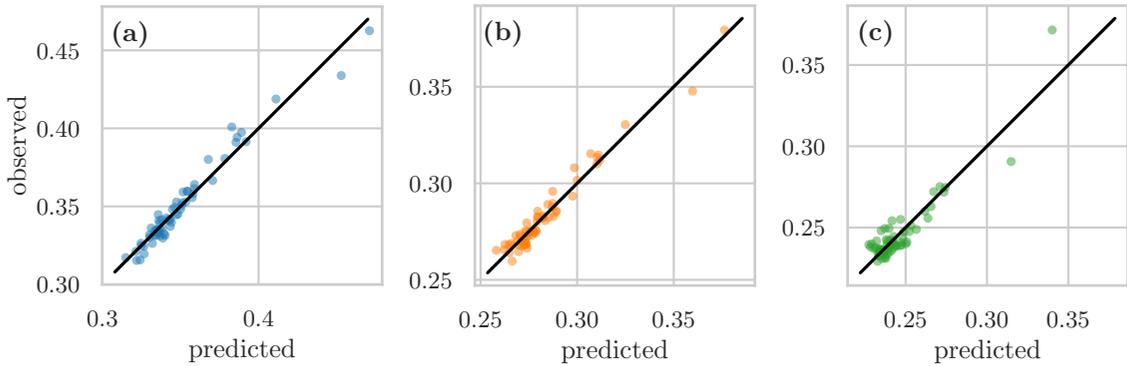}
	\caption{Comparison of predicted versus observed porosity for the linear single form models, \crefrange{eq:ss:pred_model_U1}{eq:ss:pred_model_B50}. \textbf{a} U1, \textbf{b} U7, \textbf{c} B50. The black line denotes perfect agreement.}
	\label{fig:ss:porosity_model}
\end{figure*}

The resulting linear models to predict porosity based on these two form factors were then given as
\begin{align}
n_\mathit{sf}^\mathit{U1} &= 0.019\,\mathit{EQ}^{-1} + 0.014\,\mathit{EL}^{-1} + 0.278,\label{eq:ss:pred_model_U1}\\
n_\mathit{sf}^\mathit{U7} &= 0.014\,\mathit{EQ}^{-1} + 0.007\,\mathit{EL}^{-1} + 0.234,\label{eq:ss:pred_model_U7}\\
n_\mathit{sf}^\mathit{B50} &= 0.014\,\mathit{EQ}^{-1} -0.006\,\mathit{EL}^{-1} + 0.217,\label{eq:ss:pred_model_B50}
\end{align}
where the label $\mathit{sf}$ denotes that this relation was derived for single form packings.
The model performance is visualized in \cref{fig:ss:porosity_model} by comparing measured and predicted porosity values.
According to this model, the packing's pore volume increases with the ratio $L/S$ as well as, in the unimodal cases, with the ratio $L/I$. 
From the coefficient values of the three models we can also extract that the effect of the form factors on porosity depends on the size distribution.
This effect is more pronounced for the monodisperse case U1 and gets reduced for the two cases with a wider size distribution.




\subsection{Comparison to Equivalent Ellipsoid Packings}

\begin{figure*}[t]
	\centering
	\input{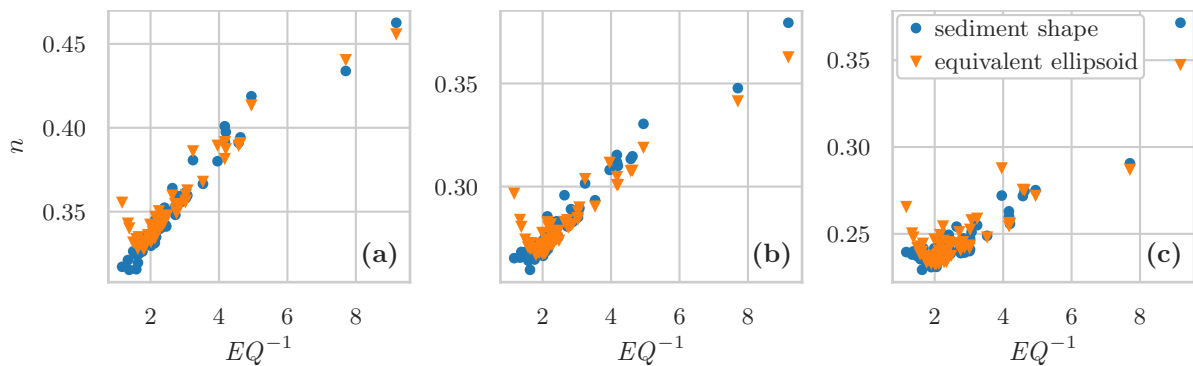}
	\caption{Comparison of porosity values obtained from packings with mesh and equivalent ellipsoids as a function of the inverse equancy. \textbf{a} U1 ($\mathit{RMSE} = 0.0087$), \textbf{b} U7 ($\mathit{RMSE} = 0.0074$), \textbf{c} B50 ($\mathit{RMSE} = 0.0074$).}
	\label{fig:ss:meshVsEllip}
\end{figure*}

We derived the form factors from equivalent ellipsoids of the distinct meshes.
This kind of form classification thus omitted all other shape-related features like the surface structure and roundness.
Thus, the question arises whether those additional effects are relevant for the observed porosity.
Since our simulation approach also allows using ellipsoidal particles, we extended our single shape analysis to packings of ellipsoids.
To this end, we carried out the same set of simulations as in the previous section but replaced the mesh-given geometry of a particle by the equivalent ellipsoid.

The comparison of porosity as a function of the inverse equancy, i.e., the dominant form factor, is shown in \cref{fig:ss:meshVsEllip}, again for the three different size distributions.
Overall, we found a very good agreement between the two types of packings, with a root-mean-square difference of around 0.008 for all three cases.
However, a clearly distinct trend was observed for inverse equancy values between 1 and 2, i.e., at the lower value range.
There, the packings using the actual sediment geometry showed a further continuous decrease of porosity, as also captured in the model from the previous section.
In contrast, the smooth ellipsoids exhibited a minimum porosity followed by a rather abrupt increase when approaching the minimal value of the inverse equancy.
This behavior is clearly visible for the unimodal cases and less pronounced for the bimodal size distribution.

From the existing studies on prolate or oblate ellipsoid packings, it is known that the porosity depends on the aspect ratio of the  ellipsoids \cite{donev2004,delaney2010,zhou2011}.
Those studies consistently reported minimal porosity values for aspect ratios of around 2, and increasing porosity values for smaller or larger aspect ratios.
Since an aspect ratio of 1 would correspond to a perfect sphere, the porosity there is the one of a random sphere packing, which in the mono-sized case is roughly 0.36 for a random packing.
Even though our equivalent ellipsoids are asymmetrical instead of prolate or oblate, the definition of the inverse equancy is actually equivalent to the aspect ratio.
Therefore, we observe the qualitatively similar behavior for the ellipsoids in \cref{fig:ss:meshVsEllip}, where the value of 0.36 is approached by the ellipsoids for case U1.
The different behavior for actual geometries and ellipsoids in this region thus has to originate from further shape-related effects that prevent this loosening of the packing and instead leads to a further compaction.
A similar effect was observed by \cite{delaney2010} for so-called superellipsoids that introduce a shape parameter $m$ to transform the shape from the smooth form of regular ellipsoids ($m=2$) to more cubical representations for larger $m$ values.
Once the blockiness of the superellipsoids was increased, the loosening effect for aspect ratios around 1 was reduced and even completely reverted for values above $m=4$.

Summarizing, we find that for the here considered sediment the actual shape, in addition to the form, has a strong influence on porosity for packings of particles with inverse equancies below 2, i.e., rather spherical particles.
Once the particle shape becomes more stretched, with $L/S > 2$, the form primarily determines porosity and other effects become less important.
Our porosity model, \crefrange{eq:ss:pred_model_U1}{eq:ss:pred_model_B50}, implicitly accounts for these additional shape-related effects by its linear form.
We, thus, assume that it is widely applicable for fluvial sediment.
However, if the particle shape significantly deviates from the here investigated ones, the model should be reevaluated for sphere-like forms, preferably with an explicit quantification of the additional shape effects.

\section{Effect of Form Distribution on the Porosity}
\label{sec:study_mixture}

In this last part, we extended our single shape studies from \cref{sec:study_single} to packings which featured a distribution of the form factors among the grains.
This way, the packings closer resembled the ones encountered in Nature.
We explored how the results obtained for single shape packings carry over to these more complex cases and how we can quantify form distributions with respect to porosity prediction. 

\subsection{Generation of Form Distribution}

\begin{figure*}[t]
	\centering
	\begin{subfigure}{0.5\textwidth}
		\input{figures/formDistStudy_I_FLvsElDistribution.pgf}
	\end{subfigure}~
	\begin{subfigure}{0.5\textwidth}
		\begin{subfigure}{0.45\textwidth}
			\begin{tikzpicture}
			\node[] (vis) at (0,0)
			{\includegraphics[width=\textwidth]{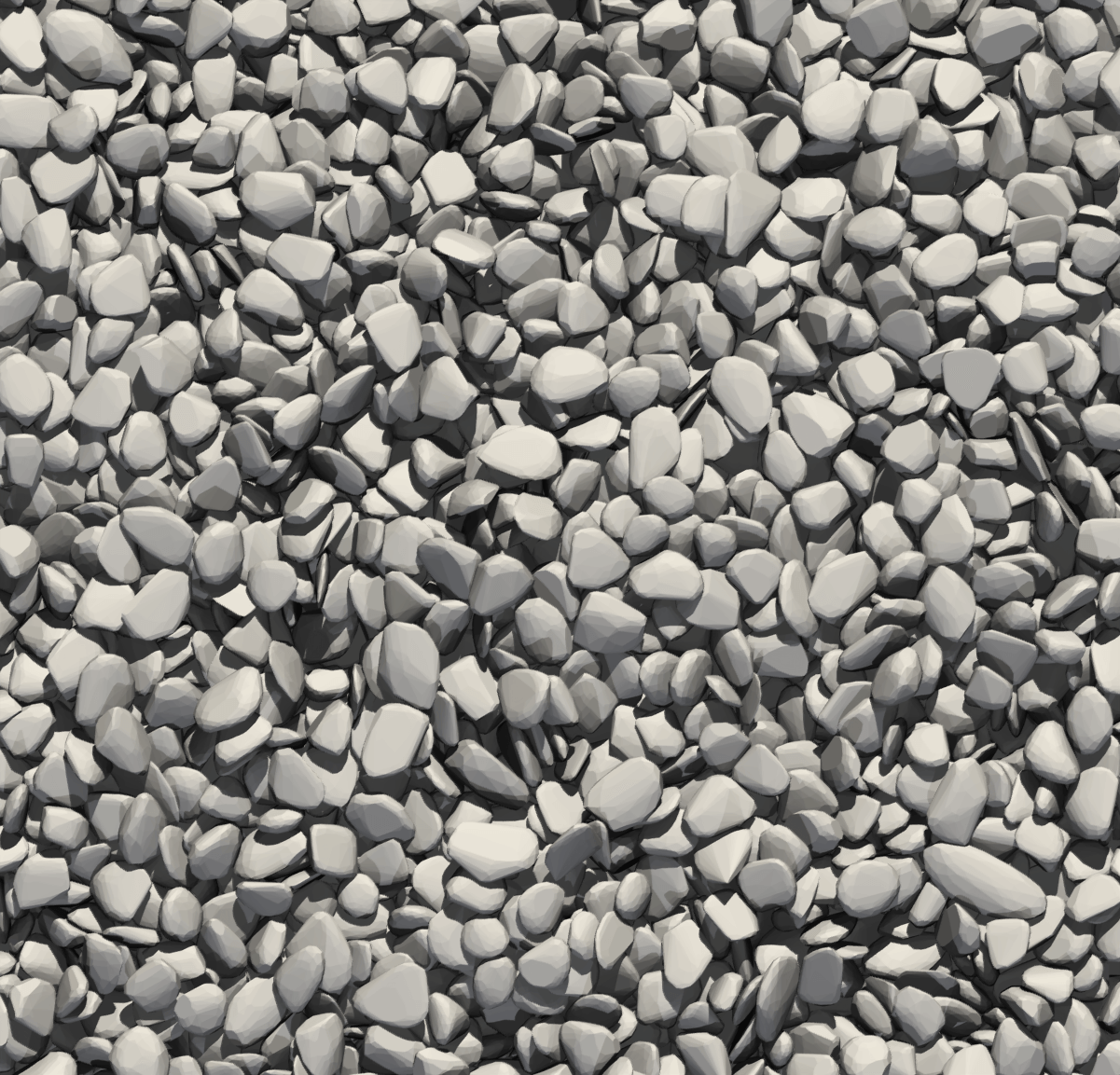}};
			\node[fill=white,anchor=north] (x) at (vis.north) {$\hat{\mu}_\mathit{FL} = 0.4, \hat{\mu}_\mathit{EL} = 0.8$};
			\end{tikzpicture}
		\end{subfigure}~
		\begin{subfigure}{0.45\textwidth}
			\begin{tikzpicture}
			\node[] (vis) at (0,0)
			{\includegraphics[width=\textwidth]{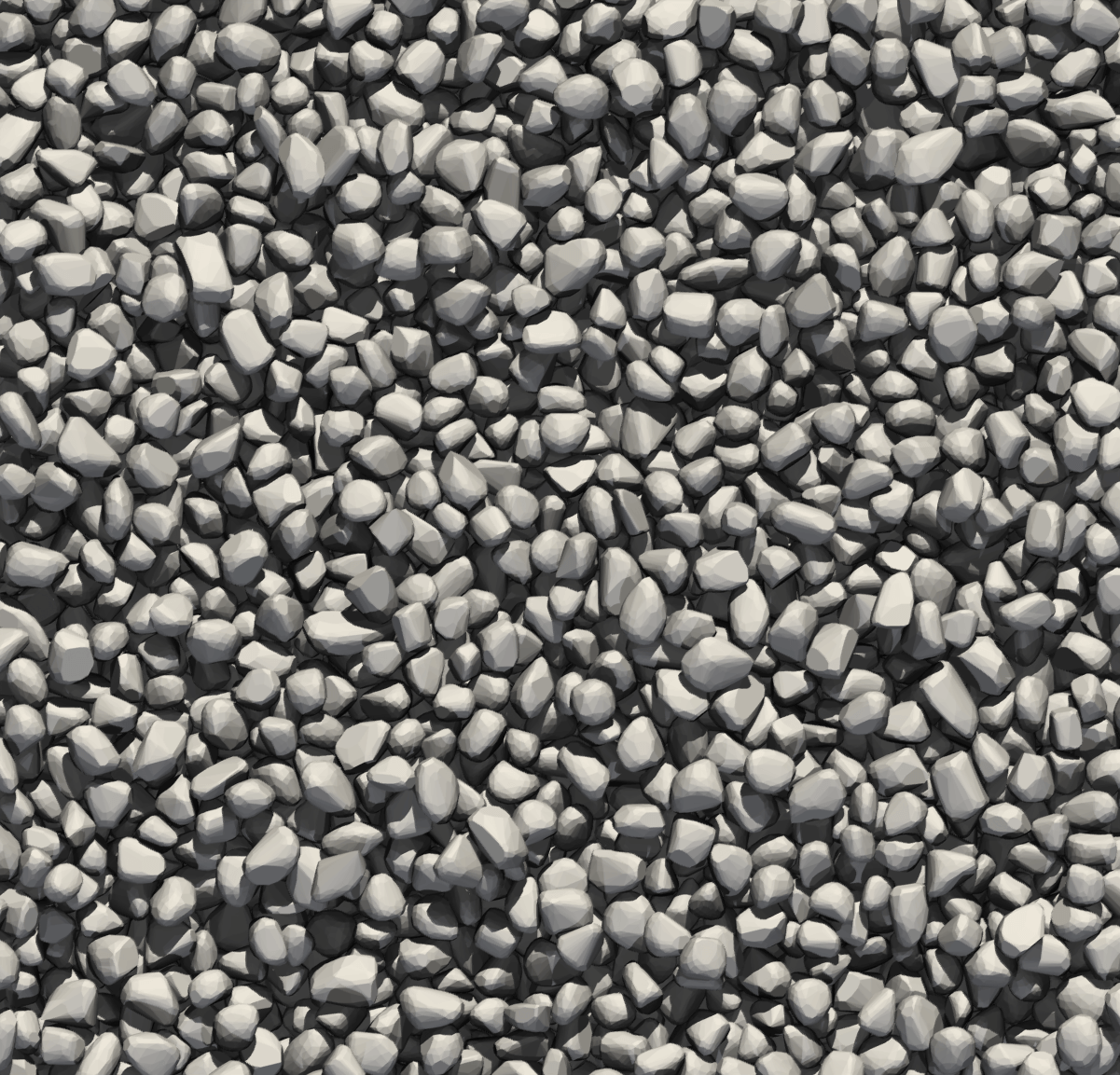}};
			\node[fill=white,anchor=north] (x) at (vis.north) {$\hat{\mu}_\mathit{FL} = 0.8, \hat{\mu}_\mathit{EL} = 0.8$};
			\end{tikzpicture}
		\end{subfigure}
		
		\begin{subfigure}{0.45\textwidth}
			\begin{tikzpicture}
			\node[] (vis) at (0,0)
			{\includegraphics[width=\textwidth]{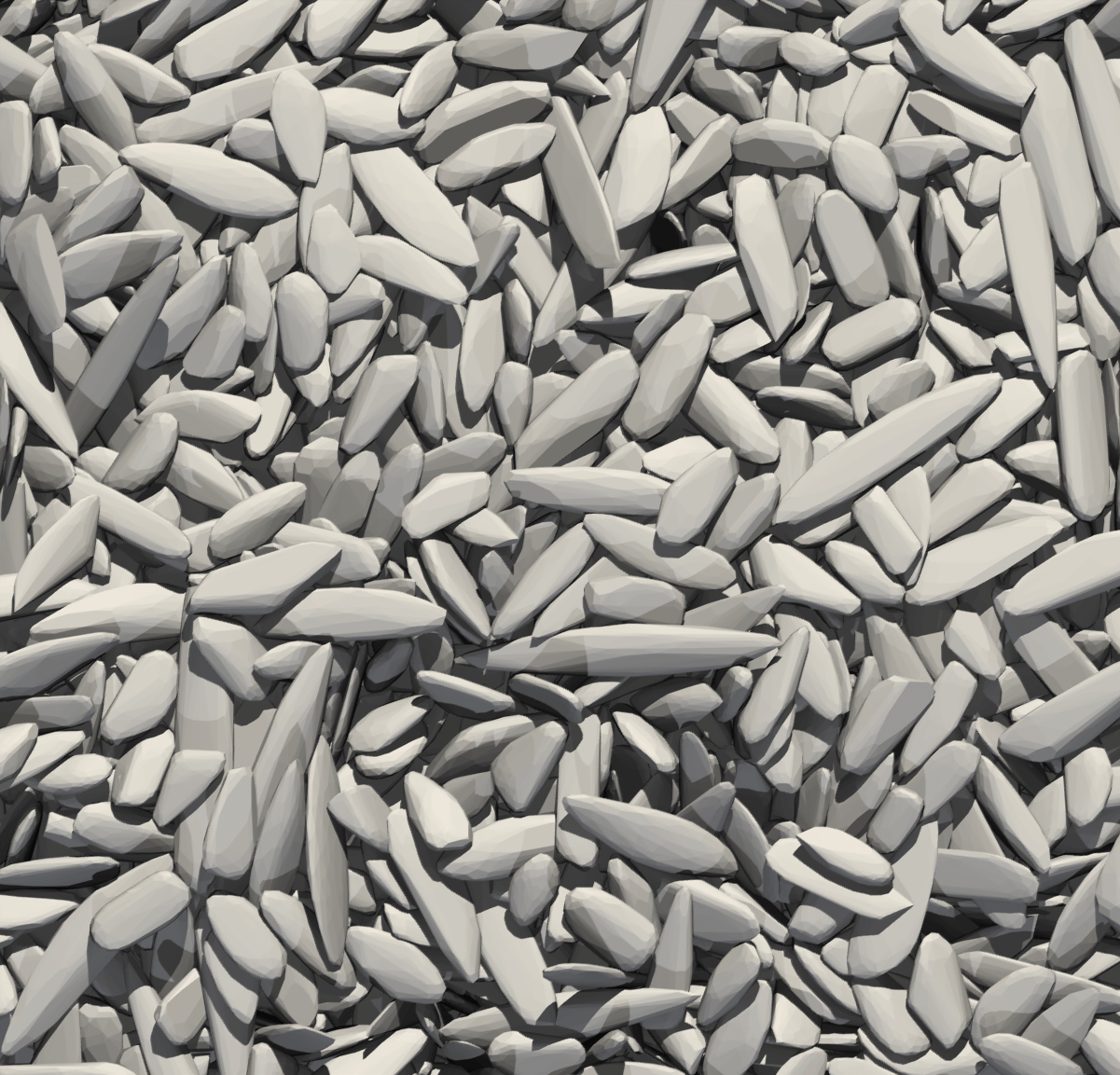}};
			\node[fill=white,anchor=south] (x) at (vis.south) {$\hat{\mu}_\mathit{FL} = 0.4, \hat{\mu}_\mathit{EL} = 0.4$};
			\end{tikzpicture}
		\end{subfigure}~
		\begin{subfigure}{0.45\textwidth}
			\begin{tikzpicture}
			\node[] (vis) at (0,0)
			{\includegraphics[width=\textwidth]{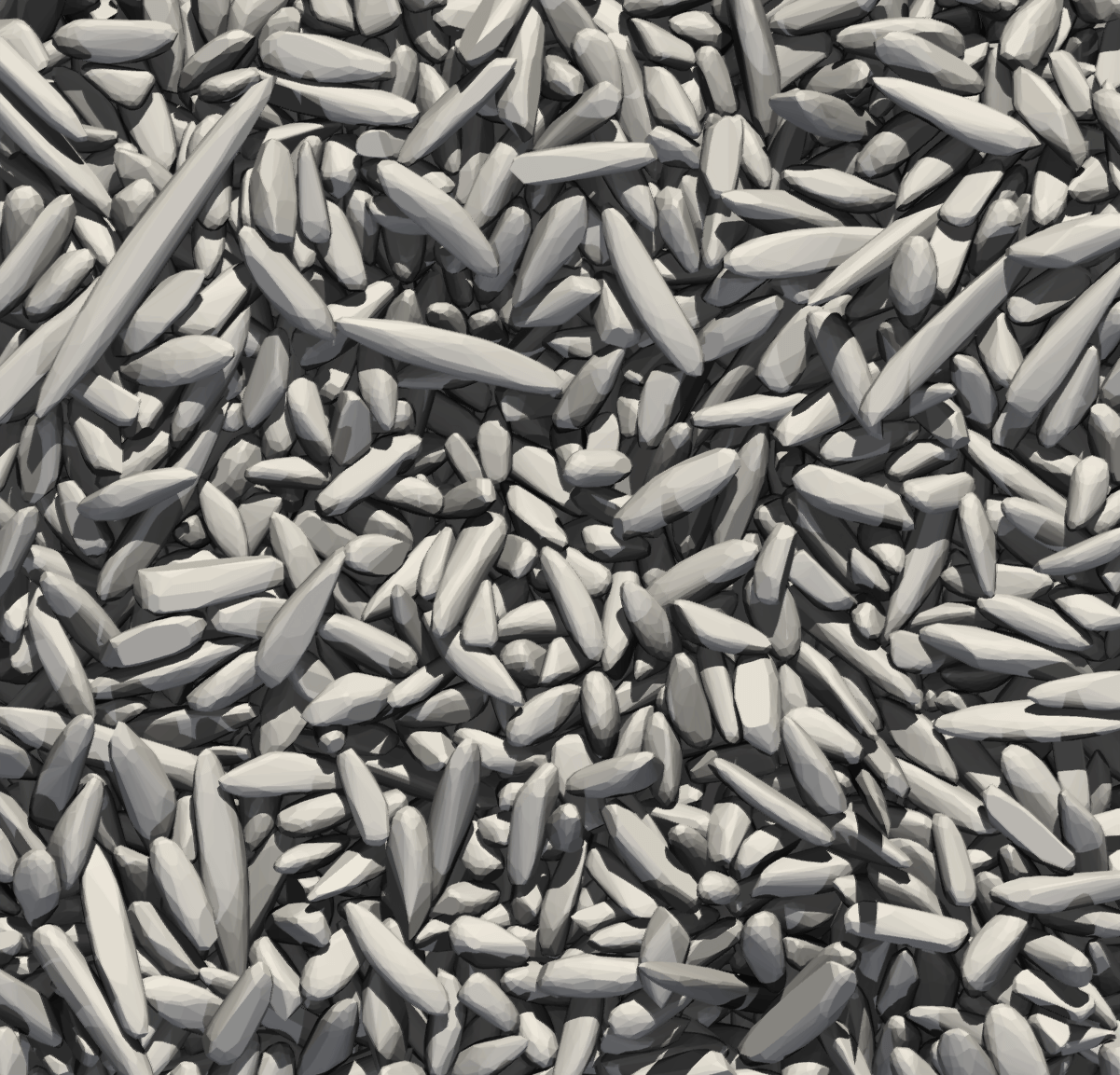}};
			\node[fill=white,anchor=south] (x) at (vis.south) {$\hat{\mu}_\mathit{FL} = 0.8, \hat{\mu}_\mathit{EL} = 0.4$};
			\end{tikzpicture}
		\end{subfigure}
	\end{subfigure}
	\caption{Visualization of the shape distributions used for the analysis for case U1. Left: Zingg diagram where, for each generated packing, the evaluated mean values of flatness and elongation are given as orange dots, while the standard deviations are displayed as the semi axes of the blue ellipses. In black, we included the parameter set of the Rhine sediment, see \cref{sec:form_meshes}. Right: Top view of four resulting packings, with the nominal mean values as given and a standard deviation of 0.1.}
	\label{fig:fm:flatness_elongation_distribution}
\end{figure*}

In order to obtain grains that exhibit a certain distribution of the form, we followed the approach outlined in \cref{sec:method:generation} to first remove the original form information from the mesh but to keep all other shape-related features.
As we have analyzed in \cref{sec:form_meshes}, the flatness and elongation of sedimentary grains can often be assumed to be normally distributed, with a number-based mean and standard deviation for both form factors.
Consequently, we sampled these two form parameters independently from two such normal distributions, $\mathcal{N}(\hat{\mu}_\mathit{FL},\hat{\sigma}_\mathit{FL})$ and $\mathcal{N}(\hat{\mu}_\mathit{EL},\hat{\sigma}_\mathit{EL})$, for each newly created particle during the generation phase of the simulation.
For our sediment geometries, the assumption of flatness and elongation being independent form parameters was verified by the insignificant correlation between these two parameters reported in \cref{fig:formFactorCorrelation}.
Then, together with the desired grain size $D$, we obtained the actual form of a grain uniquely from these parameters via:
\begin{align}
L &= \sqrt{\frac{2}{1+\textit{FL}^2}} D / \textit{EL}, \\
I &= \sqrt{\frac{2}{1+\textit{FL}^2}} D, \\
S &= \sqrt{\frac{2}{1+\textit{FL}^2}} \textit{FL}\, D.
\label{eq:fm:form_params_from_size_form}
\end{align}
To avoid unrealistically flat or elongated particles, we capped the parameters at the minimum value of 0.1. 
At the same time, to comply with the definition of flatness and elongation, the maximum admissible value was 1. 
If a value outside this range would be obtained from the distribution, we discarded it and drew another one. 

With our numerical studies, we aimed to cover a large portion of the parameter space to gain insight into the effect of form distribution.
To this end, we used two types of parameter sets to sample the relevant parameter space. 
The first involved 36 setups with all combinations when choosing the means $\hat{\mu}_\mathit{FL}$ and $\hat{\mu}_\mathit{EL}$ from the set $\{0.4, 0.6, 0.8\}$, and the standard deviations $\hat{\sigma}_\mathit{FL}$ and $\hat{\sigma}_\mathit{EL}$ from the set $\{0.1, 0.2\}$.
The second one contained 16 setups with the means from $\{0.5, 0.7\}$ and the standard deviation as before.
As shown in \cref{fig:fm:flatness_elongation_distribution}, these combinations cover the region in the Zingg diagram that appears most relevant for the present case and encompasses most findings from our literature review in \cref{sec:form_meshes}.
Since the ellipses only visualize the standard deviation, and thus the region where most grains of a packing can be found, the individual packings contained grains with a flatness-elongation combination that might be well outside of the shown area. 
Note that the depicted \textit{FL} and \textit{EL} parameters, evaluated for the generated bed, deviate slightly from the nominal ones (indicated by a hat) since the re-drawing technique for values outside the valid region introduced a shift.
To qualitatively investigate how the size distribution might affect the result, we used the cases U1 and U7 from \cref{tab:setup_experiments}.


\subsection{Porosity Model for Packings with Form Distribution}

\begin{table*}[t]
	\centering
	\begin{tabular}{cccc|cccc}
		\toprule
		\multicolumn{4}{c}{Model features} & \multicolumn{2}{|c}{U1} & \multicolumn{2}{c}{U7} \\ \cmidrule(lr){5-6}\cmidrule(lr){7-8}
		$(\mu_\mathit{EQ})^{-1}$ & $(\sigma_\mathit{EQ})^{-1}$ & $(\mu_\mathit{EL})^{-1}$ & $(\sigma_\mathit{EL})^{-1}$ & MAE & $R^2$ & MAE & $R^2$ \\ \midrule
		\checkmark & & &                                  & 6.93 & 0.939 & 5.94 & 0.867 \\
		\checkmark & \checkmark & &                       & 4.77 & 0.970 & 5.85 & 0.903 \\
		\checkmark & \checkmark & \checkmark &            & 4.26 & 0.980 & 5.96 & 0.903 \\
		\checkmark & \checkmark & \checkmark & \checkmark & 4.33 & 0.980 & 5.33 & 0.914 \\
		\checkmark & & \checkmark &                       & 6.93 & 0.949 & 6.13 & 0.868 \\
		\checkmark & \checkmark & & \checkmark            & 4.89 & 0.970 & 5.24 & 0.914 \\
		\bottomrule
	\end{tabular}
	\caption{Mean absolute errors ($\times 10^{-3}$) and coefficients of determination for the form-mixture porosity model, using different feature sets, for the size distributions U1 and U7.}
	\label{tab:fm:porosity_model_errors}
\end{table*}

Similar to \cref{sec:ss:porosity_model}, we evaluated the obtained simulation data for the respective size distribution case and computed the different form factors of \cref{tab:form_factors} and its inverse for each individual grain.
Those were then combined to determine the number-based mean and standard deviation of each of these form factors.
Additionally, we considered the inverse of each of these sample quantities as viable features.
In total, we obtained 104 features, half of which were related to the mean and the other half to the standard deviation, that were considered for the development of a predictive porosity model for packings with form mixtures.

We again first extracted the one single feature that exhibits the highest correlation with the measured porosity value.
For U1, this analysis yielded the feature $(\mu_\mathit{EQ})^{-1}$, i.e., the inverse of the mean equancy, with a correlation coefficient of 0.969.
For the wider size distribution U7, the same feature also exhibited a high correlation coefficient of 0.931.
This finding agrees well with, and thus generalizes, the single shape result from \cref{sec:ss:porosity_model} where the inverse equancy was found to be most dominant.

Further analysis along the lines of the previous single shape studies is shown in \cref{tab:fm:porosity_model_errors} which states the outcome of multivariate regression analyses with different feature sets. 
It revealed that augmenting $(\mu_\mathit{EQ})^{-1}$ with information about the standard deviation, i.e., by including $(\sigma_\mathit{EQ})^{-1}$ as a second feature, was able to reduce the error of the model predictions for both cases U1 and U7.
Adding $(\mu_\mathit{EL})^{-1}$, and thus information about a second form factor, further improved the model performance of U1, but not so for U7.
Additionally considering the standard deviation of the second form factors did not result in a better prediction for U1 but slightly improved U7.
As shown in the next-to-last row, if only information about the mean of the two form factors were supplied, the model performed hardly better than with a single form factor, and thus fell below the variants with information about the standard deviation. 
This piece of information thus seems to be a valuable ingredient for accurate porosity predictions.
For U1, it only has to be considered for one of the form factors to obtain good predictions whereas the additional standard deviation information of $\mathit{EL}$ improves the case U7.

\begin{figure*}[t]
	\centering
	\input{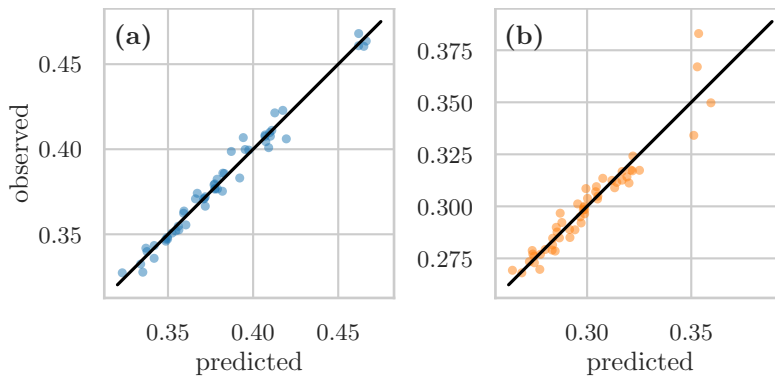}
	\caption{Comparison of predicted versus observed porosity for the linear form mixture models, \cref{eq:fm:pred_model_U1,eq:fm:pred_model_U7}. \textbf{a} U1, \textbf{b} U7. The black line denotes perfect agreement.}
	\label{fig:fm:porosity_model}
\end{figure*}

We thus decided to generally include all four features and obtained the following linear prediction model for packings with form mixtures ($\mathit{fm}$):
\begin{align}
n_\mathit{fm}^\mathit{U1} =\  & 0.032\,(\mu_\mathit{EQ})^{-1} -0.003\,(\sigma_\mathit{EQ})^{-1} + \nonumber\\
& 0.011\,(\mu_\mathit{EL})^{-1} + 0.287, \label{eq:fm:pred_model_U1}\\
n_\mathit{fm}^\mathit{U7} =\  & 0.024\,(\mu_\mathit{EQ})^{-1} -0.003\,(\sigma_\mathit{EQ})^{-1} + \nonumber\\
& 0.002\,(\mu_\mathit{EL})^{-1} + 0.001\,(\sigma_\mathit{EL})^{-1} + 0.239. \label{eq:fm:pred_model_U7}
\end{align}
The evaluation of this porosity predictor is shown in \cref{fig:fm:porosity_model}.
Similar to the single form porosity model, \cref{eq:ss:pred_model_U1,eq:ss:pred_model_U7}, the porosity is found to increase with the inverse (mean) equancy and elongation.
However, the \textit{fm} model is not a direct generalization of the \textit{sf} model, since the absence of form variance would imply $(\sigma_\mathit{EQ})^{-1} \rightarrow \infty$ and $(\sigma_\mathit{EL})^{-1} \rightarrow \infty$, resulting in an invalid porosity prediction.
To improve the generality of this model, significantly more simulations with small standard deviations would need to be added to the data set.
Instead, we here focused on the value range most relevant for real sediment, see \cref{tab:shape:flatness_elongation_literature}, for which we consider our model to be applicable.




\subsection{Discussion and Applicability} 

\begin{table*}[t]
	\centering
	\begin{tabular}{l|ccccc}
		\toprule
		& & \multicolumn{2}{c}{continuous} & \multicolumn{2}{c}{discrete}  \\ \cmidrule(lr){3-4}\cmidrule(lr){5-6}
		Case & prediction & simulation & abs. error ($\times 10^{-3}$) & simulation & abs. error ($\times 10^{-3}$)\\ \midrule
		U1 & 0.356 & 0.356 & 0.04 & 0.353 & 3.29 \\ 
		U7 & 0.285 & 0.288 & 2.74 & 0.282 & 2.72 \\ 
		\bottomrule
	\end{tabular}
	\caption{Evaluation of the porosity model, \cref{eq:fm:pred_model_U1,eq:fm:pred_model_U7}, for the case shown in \cref{fig:mesh_flatness_elongation}, i.e., with a continuous (normal) distribution of elongation and flatness or with a discrete variant, obtained by randomly sampling from the available meshes.}
	\label{tab:fm:validation}
\end{table*}

In a final step, we applied the prediction model, \cref{eq:fm:pred_model_U1,eq:fm:pred_model_U7}, for a test sample that was not contained in the data set used for the model development.
For that purpose, we simulated the cases U1 and U7 with a form distribution defined by the nominal values of the mean and standard deviations we obtained from the analysis of the available meshes (see \cref{tab:shape:flatness_elongation_literature}), i.e., $\hat{\mu}_\mathit{FL} = 0.647, \hat{\sigma}_\mathit{FL} = 0.175, \hat{\mu}_\mathit{EL} = 0.682,$ and $\hat{\sigma}_\mathit{EL}=0.139$.
Additionally, another set of simulations for both cases was carried out by randomly sampling from the available meshes as in \cref{sec:validation}.
Consequently, we kept the original form information and obtained a discrete rather than a continuous distribution of the grain form.
This setup was introduced to check the general robustness of the results when only a relatively small sample of flatness and elongation values is available, as typically the case for measurements obtained during field studies.
For both setups, the nominal mean and standard deviations of flatness and elongation were the same.
Consequently, the predicted porosity values obtained from the model were identical.

From the results in \cref{tab:fm:validation}, we see that our mixed-form porosity model yielded accurate predictions with absolute errors below 0.004 in all cases.
For U1, even an almost perfect match to packings with continuous form distributions could be observed.
The small porosity differences between the discrete and the continuous simulations were supposedly due to two reasons, in addition to the general discretization effect.
On the one hand, the actual packings in the continuous case featured slightly smaller mean and standard deviation values for flatness and elongation compared to the nominally imposed ones, due to the sampling from truncated distributions.
On the other hand, the assumption of an underlying normal distribution of both form factors in the discrete case was only approximately fulfilled, see \cref{fig:mesh_flatness_elongation}.

This validation analysis also demonstrated how the prediction model can be applied in other cases.
If shape data is available for a sufficiently large number of grains as a set of flatness - elongation pairs, as commonly displayed in the Zingg diagram, then the equancy per sample can readily be computed using $\mathit{EQ} = \mathit{EL}\, \mathit{FL}$.
The sample mean and standard deviations of \textit{EL} and \textit{EQ} are then directly available from the data as well and the predictive model can be applied.
Alternatively, shape information might only be available in terms of mean and standard deviation of flatness and elongation, see e.g. \cite{zhao2020,barrett1980}.
This was the case we assumed in the analysis above, even though we would have been able to get the model input directly from the simulation data or via the flatness and elongation information of the meshes.
Then, if one can reasonably assume that \textit{FL} and \textit{EL} are statistically independent, we can use the relations $\mu_\mathit{EQ} = \mu_\mathit{EL}\, \mu_\mathit{FL}$ and $\sigma_\mathit{EQ}^2 = (\mu_\mathit{EL}^2 + \sigma_\mathit{EL}^2)(\mu_\mathit{FL}^2 + \sigma_\mathit{FL}^2) - \mu_\mathit{EL}^2\,\mu_\mathit{FL}^2$ to obtain the input for the model, \cref{eq:fm:pred_model_U1,eq:fm:pred_model_U7}.

This step can be seen as another benefit of our rather simplistic model which does not require more complicated form factors as input where such relations might not be applicable.
With one of these two procedures we would then be able obtain porosity predictions for other sediment packings, if the actual size distribution is comparable to the ones studied here.
Note that due to geometric scaling arguments, the effect of size distributions for purely frictional-interacting systems, like sand or gravel, is to be considered relative instead of absolute.
So by implication, useful porosity predictions of any uniform packing with, e.g., a single size class can be obtained from $n_\mathit{fm}^\mathit{U1}$.

\section{Conclusion}
\label{sec:conclusion}

In this work, we investigated the influence of the sediment form on porosity of dense random deposits, clearly distinguishing between shape and form.
Generally, the form of a single grain can be described by a variety of form factors that are all combinations of only three distinct form parameters, here defined as the axes of a mass-equivalent ellipsoid.
Other, more small-scale, surface properties like angularity and roundness are thus excluded but, in combination with the form, define the shape.

We developed a numerical approach that employs friction-based interactions described by the discrete element method to simulate the settling and packing behavior of non-spherical particles.
Here, the particle shape was prescribed by a set of digitized fluvial gravel sediment grains from the Rhine river, with sizes between $2.8$ and $31.5\,$mm.
With this approach, we could accurately reproduce experimental laboratory studies in a cylindrical domain for various uni- and bimodal size distributions.
This validation thus demonstrated that such simulations can be used for predictive porosity studies.

Employing a horizontally periodic domain to represent a fraction of an actual sediment bed, we analyzed the effect of particle form on  porosity, at first imposing that all particles of the packing have the same shape and, by implication, the same form.
As a general result, we found that the form has a strong influence on porosity, especially for rather uniformly sized scenarios.
An in-depth analysis of the results revealed that porosity is primarily correlated to the inverse equancy of the grains, defined as the longest divided by the smallest form parameter.
This finding generally applies for different size distributions.
As a second form factor, that is mathematically required to uniquely describe the form, we proposed to use the inverse elongation.
Based on these two form factors, we derived a correlation that is able to accurately predict porosity.
These two form factors might thus be taken as a porosity-based classification of sediment form.

We directly compared these packings of realistic shapes to packings of equivalent ellipsoids, i.e., particles that only carry the form but no additional shape information. 
From this juxtaposition of the obtained porosity values, a good agreement was observed for inverse equancy values above two, suggesting the importance of form rather than other shape effects in cases of stretched grains.
For more spherical grains with inverse equancies below two, however, distinct behaviors were found.
There, the surface structure became more relevant and permitted denser packings than the perfectly smooth ellipsoids.
These trends also exhibited a dependence on the size distribution, where a bimodal distribution perturbed the looser packing in the ellipsoidal case.

A statistical analysis of the available surface scans, together with form data extracted from the literature, revealed that a normal distribution of the elongation and flatness of grains can often be assumed.
This observation was taken as a basis for final packing studies with form distributions, exploring the relevant parameter space via simulations.
In accordance with the single form results, we could successfully derive a porosity model based on the same two form factors, inverse equancy and inverse elongation, by considering their corresponding mean and standard deviation.
Due to the model's simplicity, we are confident that it is readily and generally applicable to other sediment packings as well, given that their size distribution is similar.

In a next step, these models have to be combined with a quantitative description of the effect of size distributions on porosity since the latter also has a strong influence on the packing behavior. 
Then, a universally applicable porosity model could be developed for sediment deposits of sand-gravel mixtures, relevant for many rivers \cite{frings2011}.
Furthermore, physical processes like a horizontally flowing fluid that might affect the deposition behavior and induce a directional structuring could be added to the simulation.
Such an extension would permit to study and quantify additional effects like imbrication or stratification, as observable in some natural sediment deposits.


\appendix
\section{Determination of Mass-Equivalent Ellipsoid}
\label{sec:app:mass_equivalent_ellipsoid}

Given a closed surface mesh that represents the geometry of a particle, we first move its center of mass into the origin.
Then, the mesh is rotated such that its principal axes of inertia are oriented along the three coordinate axes.
The inertia tensor of this mesh is then a diagonal matrix where the non-zero entries are the three moments of inertia $I_x$, $I_y$, and $I_z$.
The semi-axis of an ellipsoid with the same mass $m_p$ and moments of inertia are then given as:
\begin{align}
	a &= \sqrt{5 (-I_x + I_y + I_z) / (2m_p)} \\
	b &= \sqrt{5 (-I_y + I_x + I_z) / (2m_p)} \\
	c &= \sqrt{5 (-I_z + I_x + I_y) / (2m_p)}
\end{align}
From them, the form parameters are given as $L=2a$, $I=2b$, $S=2c$, assuming $a\geq b\geq c$.
In our case, these operations and evaluations are carried out using trimesh, a Python library \cite{trimesh}.

\section*{Acknowledgments}
The authors want to thank Axel Winterscheid for fruitful discussions that helped to shape the research.
They gratefully acknowledge the Erlangen Regional Computing Center (\url{www.rrze.fau.de/}) for funding this project by providing computing time on its supercomputers.
C.R. gratefully acknowledges the support through the German Research Foundation (DFG) grant FR 3509/4-2.

\section*{Declarations}

\textbf{Conflict of interest} The authors declare that they have no known competing financial interests or personal relationships that could have appeared to influence the work reported in this paper.

\noindent\textbf{Data availability} The simulation software, including the numerical method and the setups, is publicly available via \walberla's main repository (\url{https://walberla.net}). 
The sediment geometries of the simulative studies and the resulting data containing porosity and form factor information are available on Zenodo (\url{https://doi.org/10.5281/zenodo.6412071}).

\bibliography{Library.bib}

\end{document}